\documentclass[journal]{IEEEtran}
\usepackage{amsmath,amsfonts}

\usepackage{algorithm}
\usepackage{algorithmic}

\usepackage{array}
\usepackage{textcomp}
\usepackage{stfloats}
\usepackage{url}
\usepackage{verbatim}
\usepackage{graphicx}
\usepackage{cite}

\usepackage{amssymb}
\usepackage{xcolor}

\usepackage{comment}
\usepackage{multicol}
\usepackage{multirow}
\usepackage{subfigure}
\usepackage{setspace}
\usepackage{psfrag}
\usepackage{stmaryrd}
\usepackage{mathrsfs}
\usepackage{epstopdf}
\usepackage{enumerate}
\usepackage{booktabs}  
\usepackage{color}
\usepackage{dsfont,bm}
\usepackage{epsfig}
\usepackage{balance}

\usepackage{amsthm}

\allowdisplaybreaks[4]

\begin{document}

\title{Manifold Optimization for Hybrid Beamforming in Dual-Function Radar-Communication System}

\author{Bowen Wang,~\IEEEmembership{Graduate Student Member,~IEEE}
        Ziyang Cheng,~\IEEEmembership{Member,~IEEE}
        Zishu He,~\IEEEmembership{Senior Member,~IEEE}
\thanks{B. Wang, Z. Cheng and Z. He are with School of Information and Communication Engineering, University of Electronic Science and Technology of China, Chengdu 611731, China (B\_W\_Wang@163.com,~\{zycheng, zshe\}@uestc.edu.cn).
 \textit{(Corresponding author: Ziyang Cheng.)}
}
}



\maketitle

\begin{abstract}
We study hybrid beamforming design for millimeter wave dual-function radar communication system, which simultaneously performs downlink communication and target detection.
With two typical analog beamformer structures, we consider the hybrid beamforming design to minimize a weighted summation of radar and communication performance.  
Leveraging Riemannian optimization theory, a manifold based on the alternating direction method of multipliers is developed for the fully-connected structure. 
While for the partially-connected structure, a low-complexity Riemannian product manifold trust region algorithm is proposed to approach a near-optimal solution. 
Numerical simulations are provided to demonstrate the effectiveness of the proposed methods.
\end{abstract}


\section{Introduction}
The increasing number of wireless devices leads to more and more serious spectrum congestion.
As a consequence, there is a serious spectrum overlap between different devices, such as airbone radars and navigation systems are close to the 3.4GHz band, which partially overlap with the Long Term Evolution (LTE) systems.
Another instance is that the communication satellite (CS) systems and Wi-Fi systems operate in the C-band (4-8 GHz), which is also utilized by the military radars.
To address the spectrum congestion between radar and communication, the  communication radar spectrum sharing (CRSS) technology is a promising way. 
The existing works on the CRSS mainly focus on the two aspects.

{\it{1) Co-Existence between radar and communication systems}}

Co-existence between radar and communication systems sharing the same bandwidth has been a primary investigation field in recent years \cite{saruthirathanaworakun2012opportunistic,co-exit3,co-exit1,co-exit5,co-exit6}.
The early co-existence work focuses on dynamic spectrum access \cite{saruthirathanaworakun2012opportunistic}, where the radar or communication works when the other system is in a silent period. 
Then, benefiting from the cognitive radio development, the design of the spectrum compatible waveform with desired spectral nulls has widely been investigated \cite{co-exit1,co-exit5,co-exit6}, which provides a way to achieve co-existence for widely separated spectra systems.
However, it is not really an implementation of co-existence between radar and communication.
Note that the abovementioned works realize co-existence from radar or communication, which results in a performance loss on another side.

Recently,  the joint design scheme of radar and communication systems has become a favorable method for the co-existence systems \cite{co-exit7,co-exit8,co-design4,co-exit10,co-exit11,co-exit12}  by jointly reducing the inter-interference with each other\cite{coccia2020theory,coccia2018introduction}.
For example, the authors in \cite{co-exit7} propose a co-design of radar sampling matrix and communication codebook to co-exist between  matrix completion based MIMO radar and  MIMO communication.   
In addition,  some co-design approaches centered on communication performance are suggested in recent studies, wherein anti-radar-interference strategies are taken either at the communication receiver \cite{co-exit10}  or, using some prior information, directly at the communication transmitter \cite{co-exit11,co-exit12}.
Although these co-design methods can achieve satisfactory CRSS, the cooperation systems operating in two separated platforms need strict requirements, such as the synchronizations of frequency and sampling times. More importantly, these methods will bring extra hardware costs and complexity.

{\it{2) Dual-function radar communication systems}}

To handle the limitations of the co-existence with two platforms, the dual-function radar communication (DFRC) system, where radar and communication share in a common platform, has been regarded as a more promising approach to realize the CRSS \cite{DFRC1,DFRC4,wu2021frequency,DFRC7,blunt2010intrapulse,ciuonzo2015intrapulse,wei2022multiple,tian2022adaptive,xiao2022waveform,huang2022direct}.
Pioneering effort \cite{DFRC1} considers the single-antenna system,  in  which several integrated waveforms that can achieve both radar and communication functions are presented.  However, these methods will result in a performance loss for radar and communication because of lacking of degree of freedoms (DOFs). 
As a step further, more recent works focus on the DFRC based on the MIMO systems.
For example, a DFRC system utilizing waveform diversity in tandem with amplitude/phase control has been introduced in a number of papers \cite{DFRC4,wu2021frequency}.
Moreover, the authors in \cite{blunt2010intrapulse,ciuonzo2015intrapulse} investigate radar-embedded communications on an intrapulse, which allows for low-probability-of-intercept communication and enjoys high target detection performance \cite{ciuonzo2015intrapulse}.
To exploit the available frequency spectrum more efficiently, the authors in \cite{xiao2022waveform} investigate the DFRC waveform design in full-Duplex mode.
Additionally, a common waveform is devised to  communicate with  the downlink multi-user and detect  radar targets  in \cite{DFRC7}, in which the trade-off problem between radar and communication is established by minimize the
downlink multi-user interference (MUI) under radar-specific
constraints.

\begin{figure}[!htp]
	\centering
	\subfigure[]{
		\label{Fig_fully}
		\includegraphics[width=0.3\linewidth]{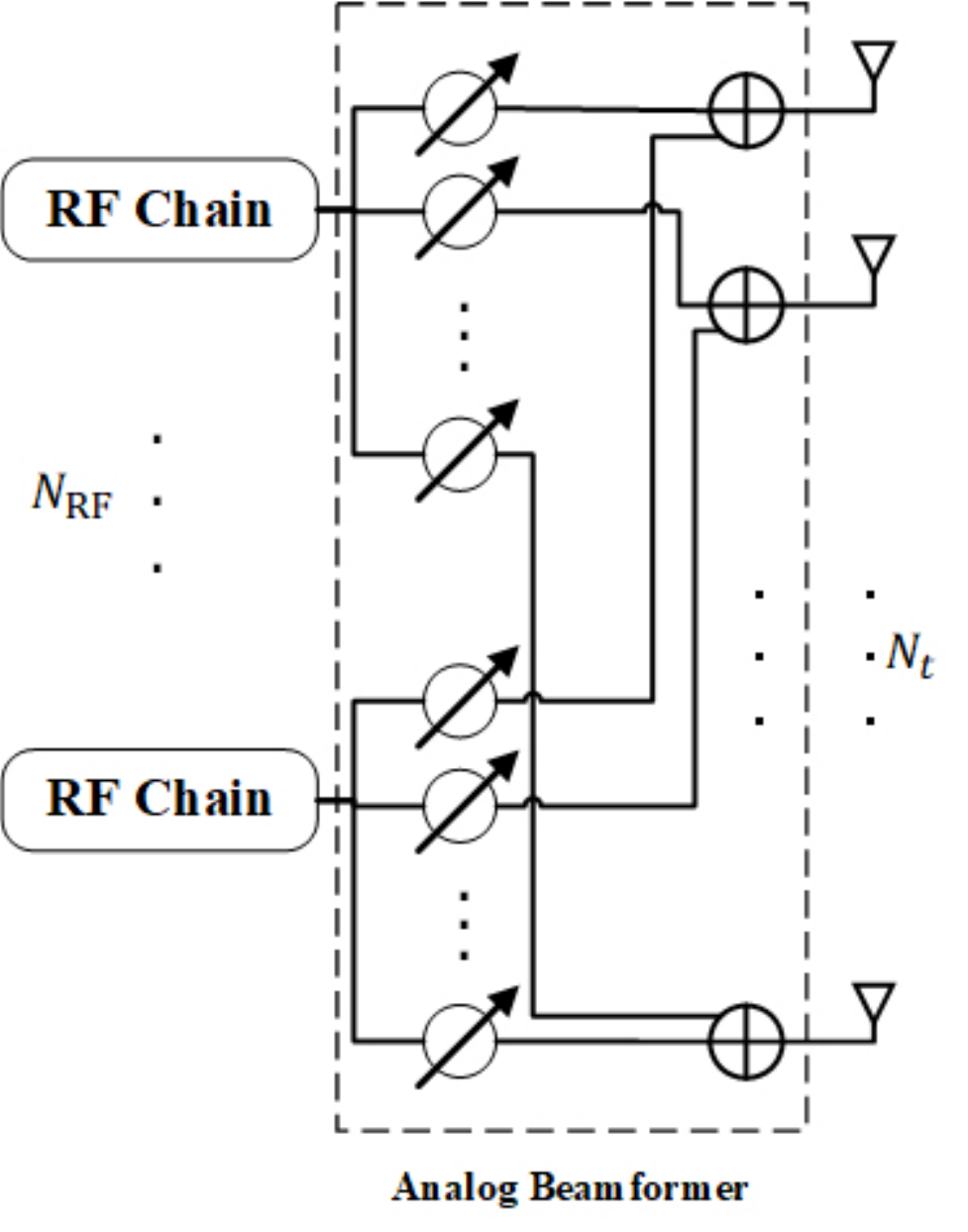}
	}
	\hspace{3em}
	\subfigure[]{
		\label{Fig_partially}
		\includegraphics[width=0.3\linewidth]{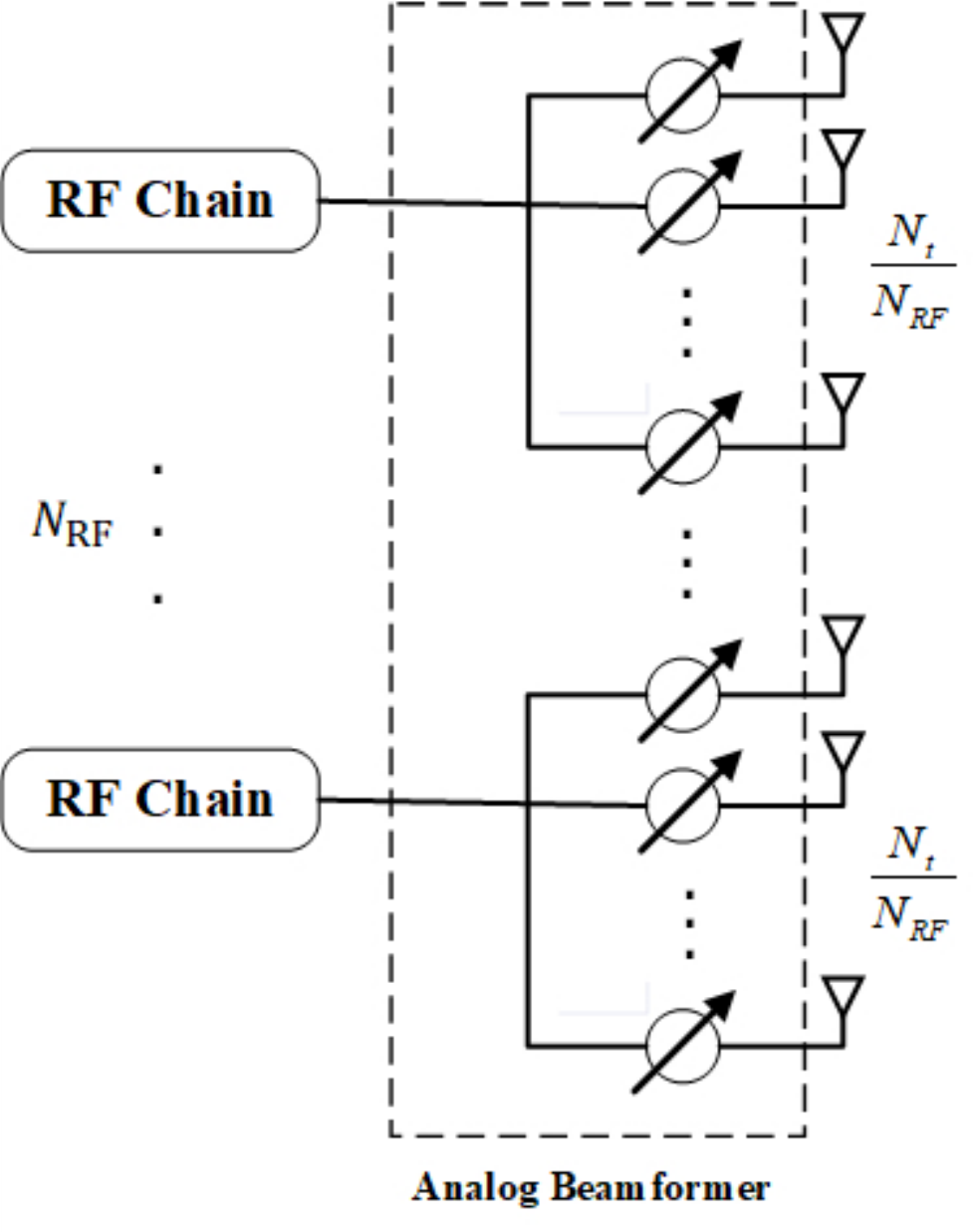}
	}
	\caption{Two structures of hybrid beamforming in mmWave MIMO systems. (a) Fully-connected structure; (b) Partially-connected structure.}
	\label{Connect_structure}
\end{figure}

It should be noted that the above methods adopt the conventional fully-digital beamforming schemes, where each antenna requires a radio frequency (RF) chain (including a Digital-to-Analog converter, up converter, etc.), leading to high hardware cost and power consumption, particularly in large-scale array systems, such as millimeter wave (mmWave) systems.
Towards system energy efficiency, hybrid beamforming (HBF) architecture consisting of analog and digital beamformers is widely studied for mmWave massive MIMO systems \cite{gao2018low,han2015large,molisch2017hybrid,ahmed2018survey,heath2016overview,HBI5}, in which a small number of RF chains is utilized to implement the digital beamformer and a large number of phase shifters (PSs) to realize the analog beamformer.
According to the mapping from the RF chain to antennas, the hybrid beamforming architectures  can be categorized into fully-connected and partially-connected (sub-connected) structures, as shown in Fig.\ref{Fig_fully} and Fig.\ref{Fig_partially}, respectively.
Compared with the fully-connected structure, the partially-connected structure, where each RF chain is connected to only a subset of antennas, is more low-cost and power efficient with slight performance loss.

It is worth pointing out that all the works above on the hybrid beamforming design focus on the communication systems.  
Recently, many researchers have devoted themselves to the hybrid beamforming design for the mmWave DFRC system \cite{DFRC8,DFRC11,DFRC10,DFRC9,Wang2022Exploiting}.
The mmWave DFRC system adopting a hybrid beamforming structure is first presented in \cite{DFRC8}, where a weighting problem is derived to facilitate the trade-off between MIMO radar and mmWave communication.
Moreover,  the authors in \cite{DFRC11} propose to jointly design the analog and digital beamformer for mmWave wideband Orthogonal Frequency Division Multiplexing (OFDM) DFRC system.
However, these works focus on designing the analog beamformer by exploiting the block coordinate descent (BCD) method, which may lead to performance loss. 
More importantly, the BCD method has a high computational complexity, which is unsuitable for the large-scale array.

More recently, manifold optimization has been vigorously developed and widely used in radar and communication designs \cite{Mani1,Mani2,Mani3,Mani4,Mani5,Mani6}. 
It converts the non-convex constraint, like constant modulus, Grassmann and Stiefel, into the corresponding manifold and performs the optimization on the manifold space \cite{Mani2}.
Moreover, many optimization solvers are based on manifold gradients, which enjoy high convergence speed and low complexity \cite{Mani1,Mani2,Mani3}.

This motivates us to study the design of analog and digital beamformers in DFRC systems via manifold minimization.
The contributions of this paper can be summarized as follows.

\begin{itemize}
	\item The problems of the hybrid beamforming design of the DFRC system with    the fully-connected and partially-connected structures are formulated by considering an optimization criterion of a weighted summation of the communication and radar beamforming errors.
	\item For the fully-connected structure, we propose the manifold alternating direction method of multipliers (MADMM) framework to tackle the hybrid beamforming optimization problem. To be more specific, we decouple the digital and analog beamformers by using the ADMM framework, such that we can optimize the digital and analog beamformers independently. Moreover, the subproblem of updating ananlog beamfomer is suboptimally solved by utilizing the Manifold optimization with a low-complexity. 
	\item For the partially-connected structure, we propose the Riemannian product manifold (RPM) optimization framework for the joint design of analog and digital beamforming. Based on the manifold optimization, the constrained optimization problem reformulate as a unconstrained optimization problem, via transforming the constant modulus and transmit power constraints into feasible search region. We derive the expressions of the Riemannian product gradient and Hessian informations, and present the trust region (TR) algorithm   to obtain the near-optimal solution.
\end{itemize}

The rest of the paper is organized as follows. 
In Section \ref{system_model}, we   introduce the system model for the mmWave DFRC system with hybrid beamforming structure. 
In Section \ref{problem_formulation}, we formulate our problem by considering the trade-off designs between radar and communication.
In Section \ref{MADMMalgo}, an efficient MADMM algorithm for the hybrid beamforming with fully-connected structure is presented.
In Section \ref{RPM-TRalgo}, we propose the Riemannian product manifold trust region algorithm (RPM-TR) to design the hybrid beamformer with partially-connected structure.
In Section \ref{Sim_Dis}, we analyze the numerical performance of the proposed algorithms and compare our approaches with the existing methods. 
Finally, we draws several conclusions in Section \ref{Conlusion}.

\textit{Notions:}Unless otherwise specified, matrices are denoted by blod uppercase letters (i.e.,${\bf{H}}$), blod lowercase letters are used for vectors (i.e.,${\bm{\alpha}}$), scalars are denoted by normal font (i.e.,$\beta$). 
${( \cdot )^T}$, ${( \cdot )^*}$ and ${( \cdot )^H}$ represent the transpose, complex conjugate and Hermitian transpose respectively. 
The set of \textit{n}-dimensional complex-valued (real-valued) vector and $N \times N$ complex-valued (real valued) matrices are denoted by ${\mathbb{C}^n}({\mathbb{R}^n})$ and $\mathbb{C}^{N \times N}\left(\mathbb{R}^{N \times N}\right)$, respectively.
$ \otimes $ and $\circ$ denote the Kronecker and Hadamard products between two matrices.
$\left| {\; \cdot \;} \right|$ and ${\left\| {\; \cdot \;} \right\|_F}$ represent determinant and Frobenius norm of argument, respectively.
${\text{vec}}( \cdot )$ and ${\text{tr}}( \cdot )$ stand for the vectorization and trace of the argument, respectively.
$\Re \mathfrak{e}( \cdot )$ denotes the real part of argument, and $\mathbb{E}$ represents expectation of a complex variable.

\section{System Model}\label{system_model}
\begin{figure}[!tb]
	\centering
	\includegraphics[width=0.9\linewidth]{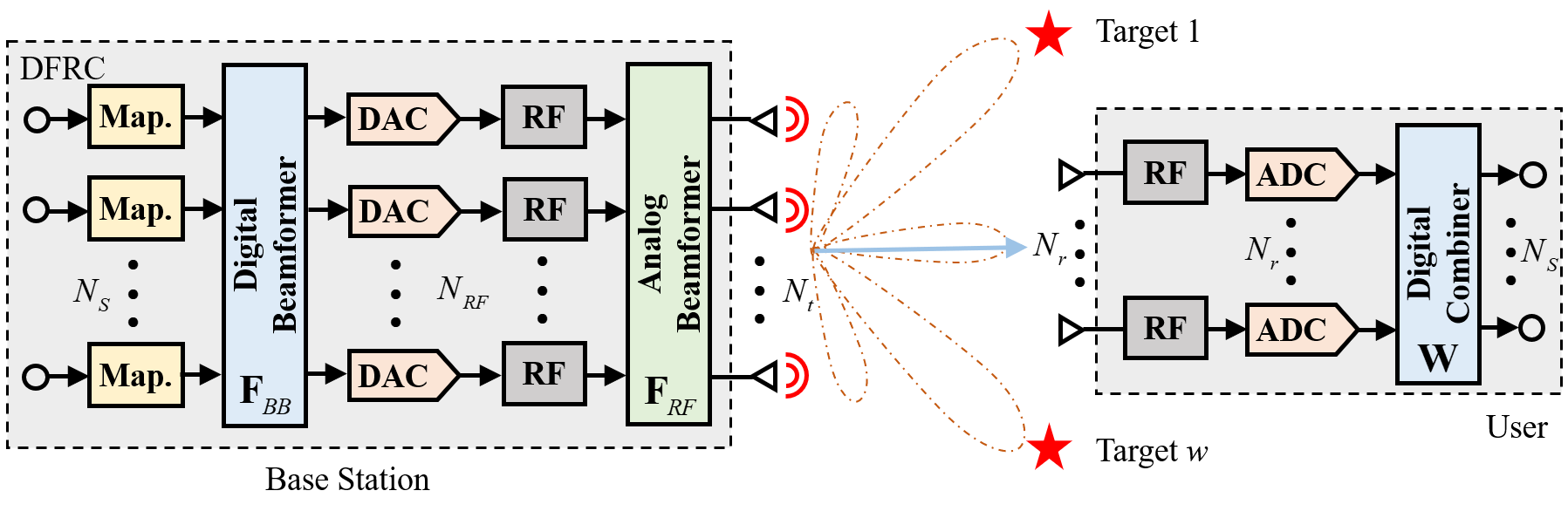}
	\caption{Overview of a mmWave dual-functional radar-communication system with hybrid analog and digital beamforming architecture at the transmitter (base station).}
	\label{fig1}
\end{figure}

Consider a mmWave MIMO-DFRC system, which  transmits probing signals to  targets  of interest and sends communication symbols to downlink user simultaneously.
As shown in Fig.\ref{fig1}, the joint transmit system is equipped with  $N_{RF}$ RF chains and $N_{t}$ antennas, which attempts to send $N_s$ data streams to a downlink user with $N_{r}$ antennas. 
Both the DFRC transmit and downlink user receive arrays are assumed to be half-wavelength spaced uniform linear arrays (ULA).
In our consideration, we assume that the number of RF chains is smaller than that of antennas at the transmit end.

\subsection{Communication Model}
In hybrid beamforming system, the data stream $\mathbf{s}$ is first processed by a digital beamformer ${{\mathbf{F}}_{BB}} \in {\mathbb{C}^{{N_{RF}} \times {N_s}}}$, and then up-converted to the carrier frequency by passing through $N_{RF}$ radio frequency (RF) chains. After that, the DFRC system uses an ${{\mathbf{F}}_{RF}} \in {\mathbb{C}^{{N_t} \times {N_{RF}}}}$ RF beamformer, which is implemented using analog phase shifters, to construct the final transmitted signal. 
Mathematically, the ${N_t} \times 1$ transmitted signal can be written as
\begin{equation}
	{{\bf x} = {{\mathbf{F}}_{RF}}{{\mathbf{F}}_{BB}}{\mathbf{s}}}
\end{equation}
where, ${\mathbf{s}} \in {\mathbb{C}^{{N_s} \times 1}}$ is the transmitted symbol vector in the baseband. Furthermore, it is assumed that $\mathbb{E}\left( {{\mathbf{s}}{{\mathbf{s}}^H}} \right) = \frac{1}{{{N_s}}}{{\mathbf{I}}_{{N_s}}}$\cite{HBI5}. The  transmitter should satisfy the transmission power constraint, i.e.,  $\left\| {{{\mathbf{F}}_{RF}}{{\mathbf{F}}_{BB}}} \right\|_F^2 = N{}_s$.

The received signal at the downlink user can be expressed as
\begin{equation}
	{{\mathbf{y}} = \sqrt \rho  {\mathbf{H}}{{\mathbf{F}}_{{{RF}}}}{{\mathbf{F}}_{{{BB}}}}{\mathbf{s}} + {\mathbf{n}}}
\end{equation}
where $\rho$ denotes the average received power, ${\mathbf{H}} \in {\mathbb{C}^{{N_r} \times {N_t}}}$ is the downlink channel matrix, and ${\mathbf{n}}$ denotes the white Gaussian noise with  zero mean and variance   $\sigma _n^2  $.

Furthermore, there are several mapping strategies form RF chains to antennas, here we consider the fully-connected and partially-connected   structures, as illustrated in Fig.\ref{Fig_fully} and Fig.\ref{Fig_partially}. 
For the fully-connected structure, the RF chains are connected to all antennas, i.e. $\left| {{{\mathbf{F}}_{RF}}\left( {i,j} \right)} \right| = 1,\forall i,j$.  While for the partially-connected structure, we assume each RF chain is connected to  ${{{N_t}} \mathord{\left/
		{\vphantom {{{N_t}} {{N_{RF}}}}} \right.
		\kern-\nulldelimiterspace} {{N_{RF}}}}$
antennas, such that we have  ${{\mathbf{F}}_{RF}} = {\text{diag}}\left[ {{{\mathbf{f}}_{R{F_1}}},{{\mathbf{f}}_{R{F_2}}},...,{{\mathbf{f}}_{R{F_{{N_{RF}}}}}}} \right]$, where ${{\mathbf{f}}_{R{F_n}}}$ is an   ${{{N_t}} \mathord{\left/
		{\vphantom {{{N_t}} {{N_{RF}}}}} \right.
		\kern-\nulldelimiterspace} {{N_{RF}}}}$ dimensional vector whose each element has a constant modulus.

To incorporate the high free-space path loss and limited scattering characteristics of the mmWave channels, we adopt a clustered channel model, i.e. Saleh-Valenzuela model \cite{gao2018low,han2015large,molisch2017hybrid,ahmed2018survey,heath2016overview,HBI5}. Under this model, the channel matrix can be written as 
\begin{equation}
	{{\mathbf{H}} = \sqrt {\frac{{{N_t}{N_r}}}{{{N_{cl}}{N_{{\text{ray}}}}}}} \sum\limits_{i = 1}^{{N_{cl}}} {\sum\limits_{l = 1}^{{N_{{\text{ray}}}}} {{\alpha _{il}}} } {{\mathbf{a}}_r}\left( {\phi _{il}^r} \right){{\mathbf{a}}_t}{\left( {\phi _{il}^t} \right)^H}}
\end{equation}
where $N_{cl}$ and $N_{ray}$ represent the number of clusters and the number of rays in each cluster, and ${{\alpha _{il}}}$ denotes the gain of $l$th ray in the $i$th propagation cluster. In addition, ${{\mathbf{a}}_r}\left( {\phi _{il}^r} \right)$ and ${{\mathbf{a}}_t}\left( {\phi _{il}^t} \right)$ represent the receive and transmit array response vectors, where ${\phi _{il}^r}$ and ${\phi _{il}^t}$ stand for azimuth angles of arrival and departure (AoAs and AoDs), respectively. 
Under the ULA assumption with half waveform length element spacing, the receive and transmit array response vectors can be expressed as
	\begin{equation}
		\begin{aligned}
			{{\bf{a}}_r}\left( \phi  \right) & = \frac{1}{{\sqrt {{N_r}} }}{\left[ {1,{e^{j\pi \sin \phi }}, \ldots ,{e^{j\pi \left( {{N_r} - 1} \right)\sin \phi }}} \right]^T},  \\
			{{\bf{a}}_t}\left( \phi  \right) & = \frac{1}{{\sqrt {{N_t}} }}{\left[ {1,{e^{j\pi \sin \phi }}, \ldots ,{e^{j\pi \left( {{N_t} - 1} \right)\sin \phi }}} \right]^T}  .
		\end{aligned}
		\label{arrayres}
	\end{equation}

In this paper, we assume that perfect channel state information (CSI) is known at the transmitter. Then, the channel's mutual information can be presented as \cite{gao2018low,han2015large,molisch2017hybrid,ahmed2018survey,heath2016overview,HBI5}
\begin{equation}
	{R = {\log _2}\left( {\left| {{{\mathbf{I}}_{{N_r}}} + \frac{\rho }{{{N_s}{\sigma _n^2}}}{\mathbf{H}}{{\mathbf{F}}_{RF}}{{\mathbf{F}}_{BB}}{\mathbf{F}}_{BB}^H{\mathbf{F}}_{RF}^H{{\mathbf{H}}^H}} \right|} \right)}
\end{equation}

\subsection{Radar Model}

In addition to downlink communication, the DFRC system aims to detect the targets of interest.
In particular, we consider a typical radar scenario, where many targets of interest exist without the signal-dependent clutter.
The baseband signal radiated towards a target located at the direction $\theta$ can be described as
\begin{equation}
	{z\left( \theta  \right) = {{\bf a}_t^H}{\left( \theta  \right)}{\mathbf{Fs}}}
	\label{eq1}
\end{equation}
where ${\mathbf{F}} = {{\mathbf{F}}_{RF}}{{\mathbf{F}}_{BB}}$ and ${\bf a}_t\left( \theta  \right)$ has the similar form as  \eqref{arrayres}.

According to \eqref{eq1}, the transmit beampattern of radar can be given as
\begin{equation}
	{P\left( \theta  \right) = \frac{1}{{{N_s}}}{{\bf a}_t}{\left( \theta  \right)^H}{\mathbf{F}}{{\mathbf{F}}^H}{{\bf a}_t}\left( \theta  \right) = \frac{1}{{{N_s}}}{{\mathbf{f}}^H}{\mathbf{Q}}\left( \theta  \right){\mathbf{f}}}
\end{equation}
where  ${\mathbf{f}} \triangleq {\text{vec}}({\mathbf{F}}) ={\text{vec}}({{\mathbf{F}}_{RF}}{{\mathbf{F}}_{BB}})$ and the matrix ${\mathbf{Q}}\left( \theta  \right)$ is given by 
\begin{equation}
	{{\mathbf{Q}}\left( \theta  \right) = {{\mathbf{I}}_{{N_s}}} \otimes \left( {{{\bf a}_t}\left( \theta  \right){{\bf a}_t}{{\left( \theta  \right)}^H}} \right)} \in {\mathbb C}^{N_t  N_s \times N_t N_s}
\end{equation}

In order to improve the detection performance, it is desired  to focus on the  transmit  power at the locations of the targets of interest  and to minimize  the power at the sidelobes.
For this purpose, we use integrated sidelobe to mainlobe ratio (ISMR) \cite{IMSR1,ISMR3} as an indicator to measure the property of the transmit beampattern, that is
\begin{equation}
	{{\text{ISMR}} = \frac{{\int_{{\Theta _s}} {P\left( \theta  \right)} {\text{d}}\theta }}{{\int_{{\Theta _m}} {P\left( \theta  \right)} {\text{d}}\theta }} = \frac{{{{\mathbf{f}}^H}{{\mathbf{L}}_s}{\mathbf{f}}}}{{{{\mathbf{f}}^H}{{\mathbf{L}}_m}{\mathbf{f}}}}}
\end{equation}
where ${\Theta _s}$ and ${\Theta _m}$ represent the sidelobe an mainlobe regions, respectively. ${\mathbf{L}}_s$ and ${\mathbf{L}}_m$ are defined as
\begin{equation}
	{{{\mathbf{L}}_s} = \int_{{\Theta _s}} {{\mathbf{Q}}\left( \theta  \right)} {\text{d}}\theta \;\;{\rm and}\;\;{{\mathbf{L}}_m} = \int_{{\Theta _m}} {{\mathbf{Q}}\left( \theta  \right)} {\text{d}}\theta }
\end{equation}

\section{Problem Formulation}\label{problem_formulation}
In this section, we first present the hybrid beamforming
design for communication-only case and  radar-only
case, respectively. Then, based on the two special cases,
we formulate  a  
weighted summation  problem  
to achieve a good   trade-off between radar and communication for the DFRC system.
\subsection{Hybrid Beamforming for Communication-Only System}
In general, communication system need to maximize the mutual  information to guarantee the quality of service (QoS). In this paper, we mainly focus on the beamformer design and assume the receiver equipping with an ideal (fully-digital) decoders. Following this strategy, the problem of the hybrid beamformer design for the communication-only system can be established as
\begin{equation}
	\begin{aligned}
		{\mathop {{\text{max}}}\limits_{{{\mathbf{F}}_{RF}},{{\mathbf{F}}_{BB}}} } & {R = {{\log }_2}\left( {\left| {{{\mathbf{I}}_{{N_r}}} + \frac{\rho }{{{N_s}{\sigma_n^2}}}{\mathbf{H}}{{\mathbf{F}}_{RF}}{{\mathbf{F}}_{BB}}{\mathbf{F}}_{BB}^H{\mathbf{F}}_{RF}^H{{\mathbf{H}}^H}} \right|} \right)} \\ 
		{{\text{s.t.}}} \quad &{\left\| {{{\mathbf{F}}_{RF}}{{\mathbf{F}}_{BB}}} \right\|_F^2 = {N_s}} \\ 
		&{\left| {{{\mathbf{F}}_{RF}}(i,j)} \right| = 1,\forall \left( {i,j} \right) \in \mathcal{A}} 
	\end{aligned}
	\label{pro1}
\end{equation}
where $\mathcal{A}  $ is the feasible set of the analog precoder. 
Due to the nonconvexity of the objective function in \eqref{pro1}, the problem is  difficult to solve numerically. Nevertheless, it has been shown in \cite{HBI5} that maximizing the objective function in \eqref{pro1} approximately leads to the minimization of the following problem: 
\begin{equation}
	\begin{aligned}
		{\mathop {{\text{min}}}\limits_{{{\mathbf{F}}_{RF}},{{\mathbf{F}}_{BB}}} } & {\left\| {{{\mathbf{F}}_{RF}}{{\mathbf{F}}_{BB}} - {{\mathbf{F}}_{Com}}} \right\|_F^2} \\ 
		{{\text{s.t.}}} \quad &{\left\| {{{\mathbf{F}}_{RF}}{{\mathbf{F}}_{BB}}} \right\|_F^2 = {N_s}} \\ 
		&{\left| {{{\mathbf{F}}_{RF}}(i,j)} \right| = 1,\forall \left( {i,j} \right) \in \mathcal{A}} 
	\end{aligned}
	\label{pro2}
\end{equation} 
where ${{{\mathbf{F}}_{Com}}}$ denotes the optimal fully digital precoder, which can be obtained by various approaches \cite{HBI5}. Here, this paper adopts zero-forcing  (ZF) method to calculated it, as  

\begin{equation}
	{{\bf{F}}_{Com}} = \frac{1}{{\sqrt {{\text{tr}}{{\left( {{\bf{H}}{{\bf{H}}^H}} \right)}^{ - 1}}} }}{{\bf{H}}^H}{\left( {{\bf{H}}{{\bf{H}}^H}} \right)^{ - 1}}
\end{equation}

\subsection{Hybrid Beamforming for Radar-Only System}
On the other hand, for the radar-only system, the transmit beamformer is desired to carefully design to attain the   minimum ISMR, the corresponding optimization problem is formulated as   
\begin{equation}
	\mathop {{\text{ min}}}\limits_{\bf{f}} {\text{ISMR}} = \frac{{{{\bf{f}}^H}{{\bf{L}}_s}{\bf{f}}}}{{{{\bf{f}}^H}{{\bf{L}}_m}{\bf{f}}}}\;\;{\text{s}}{\text{.t}}{\text{.}}\left\| {\bf{f}} \right\|_F^2 = {N_s}
	\label{pro3}
\end{equation} 
whose optimal solution can be  achieved by taking the generalized eigenvalue decomposition of $ ( {\bf L}_m, {\bf L}_s) $ \cite{IMSR1},  i.e.
\begin{equation}
	{{{\mathbf{f}}_{Rad}} = \mathcal{P}\left( {{\mathbf{L}}_s^{ - 1}{{\mathbf{L}}_m}} \right)}
\end{equation}
where the operator $ \mathcal{P}(\cdot)$ denotes the principal
eigenvector. In order to satisfy the power constraint in \eqref{pro3}, the ${{\mathbf{f}}_{Rad}}$ should be scaled as $\sqrt {{N_s}} \frac{{{{\mathbf{f}}_{Rad}}}}{{\left\| {{{\mathbf{f}}_{Rad}}} \right\|}}\xrightarrow{{}}{{\mathbf{f}}_{Rad}}$. 

Similarly to the communication-only case,  the hybrid beamformer is designed by considering the following surrogate function: 
\begin{equation}
	\begin{aligned}
		{\mathop {{\text{min}}}\limits_{{{\mathbf{F}}_{RF}},{{\mathbf{F}}_{BB}}} }&{\left\| {{{\mathbf{F}}_{RF}}{{\mathbf{F}}_{BB}} - {{\mathbf{F}}_{Rad}}} \right\|_F^2} \\ 
		{{\text{s.t.}}} \quad &{\left\| {{{\mathbf{F}}_{RF}}{{\mathbf{F}}_{BB}}} \right\|_F^2 = {N_s}} \\ 
		&{\left| {{{\mathbf{F}}_{RF}}(i,j)} \right| = 1,\forall \left( {i,j} \right) \in \mathcal{A}} 
	\end{aligned}
	\label{pro4}
\end{equation}
where ${{{\mathbf{F}}_{Rad}}}$ is the matrix form of ${{\mathbf{f}}_{Rad}}$.

\subsection{Hybrid Beamforming for DRFC System}
In DFRC system, our aim is to design the hybrid beamformer to  ensure not only the high-quality for communication, but also the detection performance for radar.  To achieve thus purpose, we introduce a weighted summation of the communication and radar beamforming errors as the optimization objective, and the resultant problem can be formulated as
\begin{subequations}
	\begin{align}
		\mathop {{\text{min}}}\limits_{{{\mathbf{F}}_{RF}},{{\mathbf{F}}_{BB}}} &\varphi \left\| {{{\mathbf{F}}_{RF}}{{\mathbf{F}}_{BB}} - {{\mathbf{F}}_{Com}}} \right\|_F^2 \nonumber \\
		& \quad + \left( {1 - \varphi } \right)\left\| {{{\mathbf{F}}_{RF}}{{\mathbf{F}}_{BB}} - {{\mathbf{F}}_{Rad}}} \right\|_F^2 \\
		{\text{s}}{\text{.t}}{\text{.}} \quad & \left\| {{{\mathbf{F}}_{RF}}{{\mathbf{F}}_{BB}}} \right\|_F^2 = {N_s} \label{pro5c}\\
		&\left| {{{\mathbf{F}}_{RF}}(i,j)} \right| = 1,\forall \left( {i,j} \right) \in \mathcal{A}   \label{pro5d}
	\end{align}
	\label{pro5}
\end{subequations}
where $\varphi  \in \left[ {0,1} \right]$ is a weighting factor (trade-off parameter) that determines the weights for communication and radar performance. 
In Sections \ref{MADMMalgo} and \ref{RPM-TRalgo}, we will consider
the RF beamformer designs for the  fully-connected structure and partially-connected structure, respectively.    
It is noticed that the problem \eqref{pro5} is difficult to tackle due to the nonconvex  constraints. 
Nevertheless, we note that the  feasible region in problem \eqref{pro5} satisfies the manifold constraints. 
Consequently, in the following, we propose to solve the nonconvex problem \eqref{pro5} with the aid of the Riemannian optimization algorithms \cite{Mani1,Mani2,Mani3}, which finds a suboptimal solution with   low-complexity.

\section{MADMM Based Hybrid Beamforming For The Fully-connected Structure}\label{MADMMalgo}
In this section, we consider the hybrid beamforming design for the fully-connected structure.  In order to obtain a near-optimal solution to  the problem \eqref{pro5} with the constraint  $\left| {{{\mathbf{F}}_{RF}}\left( {i,j} \right)} \right| = 1,\forall i,j$, the  manifold alternating direction method of multipliers (MADMM)  is proposed.

\subsection{MADMM algorithm}
As discussed in \cite{ADMM1} that solving problem \eqref{pro5}  with the ADMM framework needs to decouple   ${\mathbf{F}}_{RF}$ and ${\mathbf{F}}_{BB}$ in constraint \eqref{pro5c}. To attain such purpose,   we introduce an auxiliary variable $\mathbf{F}$, and recast the problem \eqref{pro5}   as the following problem:
\begin{equation}
	\begin{aligned}
		\mathop {{\text{min}}}\limits_{{{\bf{F}}_{RF}},{{\bf{F}}_{BB}}} & \varphi \left\| {{\bf{F}} - {{\bf{F}}_{Com}}} \right\|_F^2 + (1 - \varphi )\left\| {{\bf{F}} - {{\bf{F}}_{Rad}}} \right\|_F^2 \\
		{\text{s.t.}}\;\;\;\;\; & {\bf{F}} = {{\bf{F}}_{RF}}{{\bf{F}}_{BB}} \\
		&\left\| {\bf{F}} \right\|_F^2 = {N_S} \\
		&\left| {{{\bf{F}}_{RF}}(i,j)} \right| = 1,\forall i,j
	\end{aligned}
	\label{pro6}
\end{equation}
This will enable
us to solve the problem \eqref{pro6} more easily under the ADMM
framework. Specifically, placing  the equality constraint $  {\mathbf{F}} = {{\mathbf{F}}_{RF}}{{\mathbf{F}}_{BB}}  $ into the augmented Lagrangian function yields   
\begin{equation}
	\begin{aligned}
		\mathcal{L}\left( {{\mathbf{F}},{{\mathbf{F}}_{RF}},{{\mathbf{F}}_{BB}},{\mathbf{\Lambda }}} \right) 
		= &  \varphi \left\| {{\mathbf{F}} - {{\mathbf{F}}_{Com}}} \right\|_F^2 + (1 - \varphi )\left\| {{\mathbf{F}} - {{\mathbf{F}}_{Rad}}} \right\|_F^2 \\
		&  + \frac{\alpha }{2}\left\| {{\mathbf{F}}
			+ \frac{{\mathbf{\Lambda }}}{\alpha } - {{\mathbf{F}}_{RF}}{{\mathbf{F}}_{BB}}} \right\|_F^2 \hfill \\ 
	\end{aligned}
	\label{ADMM1}
\end{equation}

Under the ADMM framework, the solution of \eqref{pro6} can be obtained via minimizing the Lagrangian function \eqref{ADMM1}.
	Now the constraints for each variable are separated, and the coordinate descent (CD) algorithm can be utilized to solve for the variables ${\bf F}$, ${\bf F}_{RF}$ and ${\bf F}_{BB}$ iteratively.
	In the following subsection, we will discuss the solution for each block in detail.

\begin{itemize}
	\item[$\bullet$] \textbf{{Update of $\bf F$}}
\end{itemize}

For the fixed $\left\{ {{\mathbf{F}}_{RF}^{\left( {n - 1} \right)},{\mathbf{F}}_{BB}^{\left( {n - 1} \right)},{{\mathbf{\Lambda }}^{\left( {n - 1} \right)}}} \right\}$, the problem of updating the ${\bf F}$ is given by
\begin{equation}
	\begin{aligned}
		\mathop {{\text{min}}}\limits_{\bf{F}} \;\; & \varphi \left\| {{\bf{F}} - {{\bf{F}}_{Com}}} \right\|_F^2 + (1 - \varphi )\left\| {{\bf{F}} - {{\bf{F}}_{Rad}}} \right\|_F^2 \\
		& \; + \frac{\alpha }{2}\left\| {{\bf{F}} + \frac{{{{\bf{\Lambda }}^{\left( {n - 1} \right)}}}}{\alpha } - {\bf{F}}_{RF}^{\left( {n - 1} \right)}{\bf{F}}_{BB}^{\left( {n - 1} \right)}} \right\|_F^2 \\
		{\text{s}}{\text{.t}}{\text{.}}\;\;\; & \left\| {\bf{F}} \right\|_F^2 = {N_S}
	\end{aligned}
	\label{ADMMSub1}
\end{equation}

We note that its closed form solution can be attained  by using
the Karush-Kuhn-Tucker (KKT) conditions. Specifically, introducing a
Lagrange multiplier  $\Phi $ on the power constraint, we obtain the
following Lagrange function:
\begin{equation}
	\begin{aligned}
		& {\mathcal K}\left( {{\bf{F}},\Phi } \right) 
		=\varphi \left\| {{\bf{F}} - {{\bf{F}}_{Com}}} \right\|_F^2 + \left( {1 - \varphi } \right)\left\| {{\bf{F}} - {{\bf{F}}_{Rad}}} \right\|_F^2 \\
		& + \frac{\alpha }{2}\left\| {{\bf{F}} + \frac{{{{\bf{\Lambda }}^{(n - 1)}}}}{\alpha } - {\bf{F}}_{RF}^{(n - 1)}{\bf{F}}_{BB}^{(n - 1)}} \right\|_F^2 
		+ \Phi \left( {\left\| {\bf{F}} \right\|_F^2 - {N_S}} \right)
	\end{aligned}
	\label{Lag1}
\end{equation}

Based on the Lagrangian in \eqref{Lag1}, the KKT conditions for optimality are then obtained as
\begin{subequations}
	\begin{align}
		\frac{{\partial {\mathcal K}}}{{\partial {\bf{F}}}} = 2\varphi \left( {{\bf{F}} - {{\bf{F}}_{Com}}} \right) + 2\left( {1 - \varphi } \right)\left( {{\bf{F}} - {{\bf{F}}_{Rad}}} \right) \nonumber\\
		+ \alpha \left( {{\bf{F}} + \frac{{{{\bf{\Lambda }}^{(n - 1)}}}}{\alpha } - {\bf{F}}_{RF}^{(n - 1)}{\bf{F}}_{BB}^{(n - 1)}} \right) + 2\Phi {\bf{F}} &= 0 \\
		\Phi \left( {\left\| {\bf{F}} \right\|_F^2 - {N_S}} \right) &= 0 \\
		\Phi  &\geqslant 0
	\end{align}
\end{subequations}

Hence, based on the KKT conditions, the closed-form solution to ${\mathbf{F}}$ is given by
\begin{equation}
	{{{\mathbf{F}}_{opt}}{\text{ = }}\sqrt {{N_S}} \frac{{{\mathbf{\bar F}}}}{{{{\left\| {{\mathbf{\bar F}}} \right\|}_F}}}}
	\label{MADMM_so1}
\end{equation}
where ${\bf{\bar F}}{\text{  =  }}  2\varphi {{\bf{F}}_{Com}} + 2\left( {1 - \varphi } \right){{\bf{F}}_{Rad}} 
- \alpha \left( {\frac{{{{\bf{\Lambda }}^{(n - 1)}}}}{\alpha } - {\bf{F}}_{RF}^{(n - 1)}{\bf{F}}_{BB}^{(n - 1)}} \right)$.

\begin{itemize}
	\item[$\bullet$] \textbf{Update of ${\bf F}_{RF}$}
\end{itemize}

For fixed $\left\{ {{{\mathbf{F}}^{\left( n \right)}},{\mathbf{F}}_{BB}^{\left( {n - 1} \right)},{{\mathbf{\Lambda }}^{\left( {n - 1} \right)}}} \right\}$, the problem of updating the ${\bf F}_{RF}$ can be formulated as
\begin{equation}
	\begin{aligned}
		\mathop {\min }\limits_{{{\bf{F}}_{RF}}} \;\; & \left\| {{{\bf{F}}^{\left( n \right)}} + \frac{{{{\bf{\Lambda }}^{\left( {n - 1} \right)}}}}{\alpha } - {{\bf{F}}_{RF}}{\bf{F}}_{BB}^{\left( {n - 1} \right)}} \right\|_F^2  \\
		{\text{s}}.{\text{t}}. \;\;\; & \left| {{{\bf{F}}_{RF}}(i,j)} \right| = 1,\forall i,j
	\end{aligned}
	\label{ADMMSub2}
\end{equation}

Based on the manifold optimization in \cite{Mani1,Mani2,Mani3}, problem \eqref{ADMMSub2} can be converted  into the unconstrained optimization problem over manifold, given by
\begin{equation}
	{\mathop {\min }\limits_{{{\mathbf{F}}_{RF}} \in {\mathcal{M}_{{{\mathbf{F}}_{RF}}}}} {g}\left( {{{\mathbf{F}}_{RF}}} \right) = \left\| {{{\mathbf{F}}^{\left( n \right)}} + \frac{{{{\bm \Lambda} ^{(n - 1)}}}}{\alpha } - {{\mathbf{F}}_{RF}} {{\mathbf{F}}_{BB}^{(n-1)}}} \right\|_F^2}
\end{equation}
where $\mathcal{M}_{{{\mathbf{F}}_{RF}}}$ denotes  an  ${N_t}\times{N_{RF}}$ dimensional \textit{complex circle manifold} with  the geometric structure being given by
\begin{equation}
	{{\mathcal{M}_{{{\mathbf{F}}_{RF}}}} = \left\{ {{{\mathbf{F}}_{RF}} \in {\mathbb{C}^{{N_t}\times{N_{RF}}}}:\left| {{{\mathbf{F}}_{RF}}} \left(i,j\right) \right| = 1,\forall i,j} \right\}}
\end{equation}

Following \cite{Mani1,Mani2}, in order to utilize the Riemannian optimization
methods, we need to define a tangent space which is a
linear space around any point on a smooth manifold. More
concretely, the tangent space ${T_{{{\mathbf{F}}_{RF}}}}{\mathcal{M}_{{{\mathbf{F}}_{RF}}}}$ at the
point $ {\bf F}_{RF} \in {\mathcal{M}_{{{\mathbf{F}}_{RF}}}}$  is defined as
\begin{equation}
	{{T_{{{\mathbf{F}}_{RF}}}}{\mathcal{M}_{{{\mathbf{F}}_{RF}}}} = \left\{ {{\bm{\upsilon }} \in {\mathbb{C}^{{N_{RF}}\times{N_t}}}:\mathfrak{Re} \left( {{\bm{\upsilon }} \circ {\mathbf{F}}_{RF}^*}  \right)   = {\bf{0}}} \right\}}
\end{equation}
where $ \bm \upsilon$ denotes the tangent vector at point $ {\bf F}_{RF} \in {\mathcal{M}_{{{\mathbf{F}}_{RF}}}}$. 

According to the \cite{Mani1}, the Riemannian gradient of $\mathcal{M}_{{{\mathbf{F}}_{RF}}}$ 
at ${\bf F}_{\rm RF}$ is a tangent vector $ {\text{grad}}{g}({{\mathbf{F}}_{RF}})  $ given by the  projection of the Euclidean gradient $ 	{\text{Grad}}{g}({{\mathbf{F}}_{RF}})  $ on the the tangent space ${T_{{{\mathbf{F}}_{RF}}}}{\mathcal{M}_{{{\mathbf{F}}_{RF}}}}$: 
\begin{equation}
	\begin{aligned}
		{\text{grad}}g({{\bf{F}}_{RF}}) = & {\text{Pro}}{{\text{j}}_{{{\bf{F}}_{RF}}}}\left( {{\text{Grad}}g({{\bf{F}}_{RF}})} \right) \\
		= & {\text{Grad}}g({{\bf{F}}_{RF}}) 
		- \mathfrak{Re} \left( {{\text{Grad}}g{{({{\bf{F}}_{RF}})}^*} \circ {{\bf{F}}_{RF}}} \right) \circ {{\bf{F}}_{RF}}
	\end{aligned}
	\label{MADMM_grad2}
\end{equation}
where the Euclidean gradient $ 	{\text{Grad}}{g}({{\mathbf{F}}_{RF}})  $  is expressed as  
\begin{equation}
	{\text{Grad}}g({{\bf{F}}_{RF}}) = -2\left( {{{\bf{F}}^{\left( n \right)}} + \frac{{{{\bf{\Lambda }}^{\left( {n - 1} \right)}}}}{\alpha } - {{\bf{F}}_{RF}}{\bf{F}}_{BB}^{\left( {n - 1} \right)}} \right){\left( {{\bf{F}}_{BB}^{\left( {n - 1} \right)}} \right)^H}
\end{equation}

Based on the above Riemannian gradients, in what follow,  we propose an Riemannian Conjugate Gradient (RCG) algorithm, which obtains the near-optimal solution with much lower complexity \cite{Mani1}, to solve the problem \eqref{ADMMSub2}. 
For RCG method, the beamforming matrix to be updated at each iteration is computed by the descent direction and stepsize.
In the $k$-th iteration, the search direction ${\bf{\Gamma }}_{{{\bf{F}}_{RF}}}^{\left( k \right)}$ is determined by the Riemannian gradient ${\text{grad}}g({\bf{F}}_{RF}^{\left( k \right)})$ and the $\left(k-1\right)$-th search direction ${\bf{\Gamma }}_{{{\bf{F}}_{RF}}}^{\left( k-1 \right)}$, which is given by
\begin{equation}
	{\bf{\Gamma }}_{{{\bf{F}}_{RF}}}^{\left( k \right)} =  - {\text{grad}}g\left( {{\bf{F}}_{RF}^{\left( k \right)}} \right) + \sigma _{{{\bf{F}}_{RF}}}^{\left( {k - 1} \right)}{\text{Pro}}{{\text{j}}_{{\bf{F}}_{RF}^{\left( k \right)}}}\left( {{\bf{\Gamma }}_{{{\bf{F}}_{RF}}}^{\left( {k - 1} \right)}} \right)
	\label{RCGdire2}
\end{equation}
where   ${\sigma _{{{\mathbf{F}}_{RF}}}}$ is the polak-Ribiere parameter \cite{Mani5}, defined as 
\begin{equation}
	{\sigma _{{{\mathbf{F}}_{RF}}}^{\left( k \right)} = \frac{{\left\langle {{\text{grad}}g({\mathbf{F}}_{RF}^{\left( k \right)}),{\bm{\chi}} _{{{\mathbf{F}}_{RF}}}^{\left( k \right)}} \right\rangle }}{{\left\langle {{\text{grad}}g({\mathbf{F}}_{RF}^{\left( {k - 1} \right)}),{\text{grad}}g({\mathbf{F}}_{RF}^{\left( {k - 1} \right)})} \right\rangle }}}
	\label{RCGPR2}
\end{equation}
with  $\left\langle { \cdot , \cdot } \right\rangle$ denoting the usual Euclidean inner product, i.e.,  $\left\langle {{\bf{P}},{\bf{Q}}} \right\rangle  = \mathfrak{Re} \left( \text {tr}\left(  {{{\bf{P}}^H}{\bf{Q}}} \right) \right)$, and ${{\bm{\chi}} _{{{\mathbf{F}}_{RF}}}}$ standing for the difference between the current and the previous gradients, and it is expressed as
\begin{equation}
	{\bm{\chi}}_{{{\bf{F}}_{RF}}}^{\left( k \right)} = {\text{grad}}{g}({\bf{F}}_{RF}^{\left( k \right)}) - {\text{Pro}}{{\text{j}}_{{\bf{F}}_{RF}^{\left( {k - 1} \right)}}}\left( {{\text{grad}}{g}({\bf{F}}_{RF}^{\left( {k - 1} \right)})} \right)
\end{equation}

Then, the stepsize $\mu _{{{\bf{F}}_{RF}}}^{\left( {k} \right)}$ is chosen by the Armijo backtracking line search method \cite{Mani5}, and the beamforming matrix ${\mathbf{F}}_{RF}^{\left( k \right)}$ to be updated at the step $\left(k+1\right)$ is calculated by the \textit{retraction} on the manifold, which is
\begin{equation}
	{\bf{F}}_{RF}^{\left( k+1 \right)} = {{\mathcal R}_{{\bf{F}}_{RF}^{\left( {k} \right)}}}\left( {\mu _{{{\bf{F}}_{RF}}}^{\left( {k} \right)}{\bf{\Gamma }}_{{{\bf{F}}_{RF}}}^{\left( {k } \right)}} \right)
	\label{RCGrecter2}
\end{equation}
where $\mathcal{R}$ is a mapping operator, named as retraction, which  maps a vector from the tangent space onto the manifold itself,  given by
\begin{equation}
	\begin{aligned}
		& {{\mathcal R}_{{{\bf{F}}_{RF}}}} \left( {{\mu _{{{\bf{F}}_{RF}}}}{{\bf{\Gamma }}_{{{\bf{F}}_{RF}}}}} \right) \\ 
		& \quad = \left( {{{\bf{F}}_{RF}} + {\mu _{{{\bf{F}}_{RF}}}}{{\bf{\Gamma }}_{{{\bf{F}}_{RF}}}}} \right) \circ \frac{{\text{1}}}{{\left| {\left( {{{\bf{F}}_{RF}} + {\mu _{{{\bf{F}}_{RF}}}}{{\bf{\Gamma }}_{{{\bf{F}}_{RF}}}}} \right)} \right|}}
	\label{Retractions}
	\end{aligned}
\end{equation}

Based on the above analysis, the RCG Algorithm for solving \eqref{ADMMSub2} is summarized by Algorithm \ref{alg2}.
\begin{algorithm}[htb] 
	\caption{RCG Algorithm for solving \eqref{ADMMSub2}} 
	\label{alg2} 
	\begin{algorithmic}[1] 
		\REQUIRE Set the initial variables ${\mathbf{F}}^{\left( {n} \right)}$, ${\mathbf{F}}_{BB}^{\left( {n - 1} \right)}$, 
		${{\mathbf{\Lambda }}^{\left( {n - 1} \right)}}$, ${{\mathbf{F}}_{Rad}}$, 
		${{\mathbf{F}}_{Com}}$, $\alpha$ and $\eta $ and ${K_{\max }}$.
		\ENSURE ${\mathbf{F}}_{RF}^{(opt)}$.
		\STATE \textbf{Initialization:} Randomly set ${{\mathbf{F}}_{{RF}}^{\left( 0 \right)}} \in {\mathcal{M}_{{\mathbf{F}}_{RF}}}$
		\STATE Set $k = 1$.
		\WHILE{$k \leqslant {K_{\max }}$ and ${\left\| {{\text{grad}}{g}\left( {{\mathbf{F}_{RF}^{\left( k \right)}}} \right)} \right\|_F} \geqslant \eta $}
		\STATE Compute the stepsize $\mu _{{{\bf{F}}_{RF}}}^{\left( {k} \right)}$ using the Armijo Rule.
		\STATE Calculate the Polak-Ribiere parameter based on \eqref{RCGPR2}.
		\STATE Obtain the descent direction according to \eqref{RCGdire2}.
		\STATE Update the beamforming matrix ${\mathbf{F}_{RF}^{\left( {k + 1} \right)}}$ by \eqref{RCGrecter2}.
		\STATE $k = k + 1$	
		\ENDWHILE
		\STATE ${\mathbf{F}}_{RF}^{(opt)} =  {{\mathbf{F}}_{RF}^{\left(k\right)}} $
	\end{algorithmic}
\end{algorithm}

\begin{itemize}
	\item[$\bullet$] \textbf{Update of ${\bf F}_{BB}$}
\end{itemize}

For fixed $\left\{ {{{\mathbf{F}}^{\left( n \right)}},{\mathbf{F}}_{RF}^{\left( n \right)},{{\mathbf{\Lambda }}^{\left( {n - 1} \right)}}} \right\}$, the problem of updating the ${\bf F}_{BB}$ can be expressed as
\begin{equation}
	{\mathop {\min }\limits_{{{\mathbf{F}}_{BB}}}    \left\| {{{\mathbf{F}}^{\left( n \right)}} + \frac{{{{\mathbf{\Lambda }}^{\left( {n - 1} \right)}}}}{\alpha } - {\mathbf{F}}_{RF}^{\left( n \right)}{{\mathbf{F}}_{BB}}} \right\|_F^2}
	\label{ADMMS2}
\end{equation}

It is easy to observe that the problem \eqref{ADMMS2} is convex Quadratic Program without constraint, and its closed-form solution can be attained as
\begin{equation}
	{\bf{F}}_{BB}^{\left( n \right)} = {\left( {{{\left( {{\bf{F}}_{RF}^{\left( n \right)}} \right)}^H}{\bf{F}}_{RF}^{\left( n \right)}} \right)^{ - 1}}{\left( {{\bf{F}}_{RF}^{\left( n \right)}} \right)^H}\left( {{{\bf{F}}^{\left( n \right)}} + \frac{{{{\bf{\Lambda }}^{\left( {n - 1} \right)}}}}{\alpha }} \right)
	\label{MADMM_so2}
\end{equation}

\begin{itemize}
	\item[$\bullet$] \textbf{Update Penalty Parameter $\alpha$}
\end{itemize}

In conventional ADMM algorithm \cite{ADMM1}, using  a fixed penalty parameter may lead to  poor convergence performance. In order to improve the convergence and make performance less dependent on the initial choice of the penalty parameter, we use the following update scheme of   $\alpha$\cite{ADMM1}:
\begin{equation}
	{{\alpha ^{\left( n \right)}} = \left\{ {\begin{array}{*{20}{l}}
				{\beta {\alpha ^{\left( {n - 1} \right)}},\;\;\;\;{\text{if}}\;\frac{{\left\| {{{\mathbf{F}}^{(n)}} - {\mathbf{F}}_{RF}^{(n)}{\mathbf{F}}_{BB}^{(n)}} \right\|_F^2}}{{\left\| {{{\mathbf{\Lambda }}^{(n)}} - {{\mathbf{\Lambda }}^{(n - 1)}}} \right\|}} > \gamma } \\ 
				{{{{\alpha ^{\left( {n - 1} \right)}}} \mathord{\left/
							{\vphantom {{{\alpha ^{\left( {n - 1} \right)}}} \beta }} \right.
							\kern-\nulldelimiterspace} \beta },\;\;{\text{if}}\;\frac{{\left\| {{{\mathbf{F}}^{(n)}} - {\mathbf{F}}_{RF}^{(n)}{\mathbf{F}}_{BB}^{(n)}} \right\|_F^2}}{{\left\| {{{\mathbf{\Lambda }}^{(n)}} - {{\mathbf{\Lambda }}^{(n - 1)}}} \right\|}} < \frac{1}{\gamma }} \\ 
				{{\alpha ^{\left( {n - 1} \right)}},\;\;\;\;\;\;{\text{otherwise}}} 
		\end{array}} \right.}
	\label{uppara}
\end{equation}
where $\beta  > 1$ and $\gamma  > 1$. In our case, we set $\beta  = 2$ and $\gamma  = 10$.

According to the above analysis, the proposed method for designing the hybrid beamforming in the case of fully-connected structure  is summarized in Algorithm \ref{alg4}.

\begin{algorithm}[htb] 
	\caption{MADMM for fully-connected structure hybrid beamforming design} 
	\label{alg4} 
	\begin{algorithmic}[1] 
		\REQUIRE Set the initial variables ${{\mathbf{F}}^0}$, ${\mathbf{F}}_{RF}^0$, ${\mathbf{F}}_{BB}^0$, ${{\mathbf{\Lambda }}^0}$, $\alpha^0$, ${\bf{F}}_{Rad}$, ${\bf{F}}_{Com}$ and $N_{max}$.
		\ENSURE ${\bf{F}}_{RF}^{\left( {opt} \right)}$ and ${\bf{F}}_{BB}^{\left( {opt} \right)}$.
		\STATE \textbf{Initialization:} Set $n = 1$, $\beta=2$ and $\gamma=10$.
		\WHILE{ $n \leqslant {N_{\max }}$}
		\STATE  For given ${{\mathbf{F}}_{RF}^{\left( {n - 1} \right)},{\mathbf{F}}_{BB}^{\left( {n - 1} \right)},{{\mathbf{\Lambda }}^{\left( {n - 1} \right)}}}$, obtain the solution of ${{\mathbf{F}}^{\left( n \right)}}$ by using \eqref{MADMM_so1}
		\STATE  For given ${{\mathbf{F}}^{\left( n \right)}},{\mathbf{F}}_{BB}^{\left( {n - 1} \right)},{{\mathbf{\Lambda }}^{\left( {n - 1} \right)}}$, obtain the solution of ${\mathbf{F}}_{RF}^{\left( n \right)}$ according to Algorithm \ref{alg2}.
		\STATE  For given ${{\mathbf{F}}^{\left( n \right)}},{\mathbf{F}}_{RF}^{\left( n \right)},{{\mathbf{\Lambda }}^{\left( {n - 1} \right)}}$, update ${\mathbf{F}}_{BB}^{\left( n \right)}$ using \eqref{MADMM_so2}.
		\STATE ${{\bf{\Lambda }}^{\left( n \right)}} = {{\bf{\Lambda }}^{\left( {n - 1} \right)}} + \alpha \left( {{{\bf{F}}^{\left( n \right)}} - {\bf{F}}_{RF}^{\left( n \right)}{\bf{F}}_{BB}^{\left( n \right)}} \right)$.
		\STATE  Update $\alpha^{\left( n \right)}$ according to \eqref{uppara}.
		\STATE  $n = n + 1$
		\ENDWHILE
		\STATE ${\bf{F}}_{RF}^{\left( {opt} \right)} = {\bf{F}}_{BB}^{\left( {{N_{max}}} \right)}$ and ${\bf{F}}_{BB}^{\left( {opt} \right)} = {\bf{F}}_{BB}^{\left( {{N_{max}}} \right)}$
	\end{algorithmic} 
\end{algorithm}

At the end of this section, we analyze the complexity of the proposed MADMM algorithm  for the hybrid beamforming in the case of fully-connected structure.
It should be noted that the main computational complexity of the MADMM method is caused by solving three subproblems, i.e. Updating ${\bf F}$, ${\bf F}_{RF}$ and ${\bf F}_{BB}$.
Updating ${\bf F}$ needs complexities of $\mathcal{O}\left( {{N_s}{N_{RF}}{N_t}} \right)$.
Using the RCG method updates ${\bf F}_{RF}$ with the complexities of  $\mathcal{O}\left( {{I_1}{N_s}{N_{RF}}N_t} \right)$, where ${{I_1}}$ are the number of the iterations required in RCG methods.
Updating ${\bf F}_{BB}$ needs complexities of $\mathcal{O}\left( {N_{RF}^2{N_t} + {N_s}{N_{RF}}{N_t}} \right)$.
To summarize, the overall complexity of the MADMM method is ${\mathcal O}\left( {{I_0}{I_1}{N_s}{N_{RF}}{N_t}} \right)$, where $I_0$ is the number of  MADMM iterations.

\section{Riemannian Product Manifold Based Hybrid Beamforming For The Partially-connected Structure}\label{RPM-TRalgo}
In contract to  the fully-connected structure, the partially-connected structure adopts less phase shifters and has lower power and hardware cost 
in mmWave systems.
However, due to the fact that the analog beamfomer matrix is a block diagonal structure with the unit modulus,  the corresponding optimization problem is hard to tackle.
Therefore, in this section, we propose the Riemannian product manifold trust region (RPM-TR) algorithm to design the hybrid beamfoming with  low-complexity.

\subsection{Problem Reformulation}
Due to the special structure of the ${{{\mathbf{F}}_{RF}}}$, we have the following   equality: 
\begin{equation}
	{\left\| {{{\mathbf{F}}_{RF}}{{\mathbf{F}}_{BB}}} \right\|_F^2 = \frac{{{N_t}}}{{{N_{RF}}}}\left\| {{{\mathbf{F}}_{BB}}} \right\|_F^2 = {N_S}}
\end{equation}

We note that the relationship between the  partially-connected structure  and  fully-connected  can be established as follows,  
\begin{equation}
	{{{\mathbf{F}}_{{ {RF}}}} = {{\mathbf{F}}_{PS}} \circ {{\mathbf{F}}_D}}
\end{equation}
where ${{\mathbf{F}}_{PS}} \in {\mathbb{C}^{{N_t} \times {N_{RF}}}}$ stands for the fully-connected  phase shifter matrix, i.e.  $\left| {{{\mathbf{F}}_{PS}}\left( {i,j} \right)} \right| = 1,\forall i,j$, and ${{\mathbf{F}}_D} \in {\mathbb{C}^{{N_t} \times {N_{RF}}}}$ denotes 0-1 connection-state matrix.
For the partially-connected structure, the connection-state matrix is ${{\mathbf{F}}_D} = {\text{diag}}\left( {{{\mathbf{1}}_z},{{\mathbf{1}}_z},...,{{\mathbf{1}}_z}} \right) \in {\mathbb{C}^{{N_t} \times {N_{RF}}}}$, where $z = \frac{{{N_t}}}{{{N_{RF}}}}$
Based on the above analysis, the problem \eqref{pro5} can be rewritten as 
\begin{equation}
	\begin{aligned}
		\mathop {{\text{min}}}\limits_{{{\bf{F}}_{PS}},{{\bf{F}}_{BB}}} \;\; & \varphi  \left\| {\left( {{{\bf{F}}_{PS}} \circ {{\bf{F}}_D}} \right){{\bf{F}}_{BB}} - {{\bf{F}}_{Com}}} \right\|_F^2 \\
		& \;\; + (1 - \varphi )\left\| {\left( {{{\bf{F}}_{PS}} \circ {{\bf{F}}_D}} \right){{\bf{F}}_{BB}} - {{\bf{F}}_{Rad}}} \right\|_F^2 \\
		{\text{s}}{\text{.t}}{\text{.}} \;\;\;\;\;\; & \left\| {{{\bf{F}}_{BB}}} \right\|_F^2 = \frac{{{N_S}{N_{RF}}}}{{{N_t}}}, \\
		&\left| {{{\bf{F}}_{PS}}}\left( i,j \right)  \right|_{} = 1,\forall i,j 
	\end{aligned}
	\label{RPM1}
\end{equation}

For the convenience of deriving the problem, we use the  $J\left( {{{\mathbf{F}}_{PS}},{{\mathbf{F}}_{BB}}} \right)$ denoting the objective function in \eqref{RPM1}.

\subsection{Geometric Structure of the Riemannian Product Manifold}
In order to solve this problem effectively, we recast the constrained optimization problem \eqref{RPM1} as an standard optimization form over a product manifold $\mathcal{M}$, as 
\begin{equation}
	{\mathop {{\text{min}}}\limits_{\left( {{{\mathbf{F}}_{PS}},{{\mathbf{F}}_{BB}}} \right) \in \mathcal{M}} J\left( {{{\mathbf{F}}_{PS}},{{\mathbf{F}}_{BB}}} \right)}
	\label{RPMF1}
\end{equation}
where $\mathcal{M} = {\mathcal{M}_{{{\mathbf{F}}_{PS}}}} \times {\mathcal{M}_{{{\mathbf{F}}_{BB}}}}$ denotes the product manifold of ${\mathcal{M}_{{{\mathbf{F}}_{PS}}}} = \left\{ {{{\mathbf{F}}_{PS}} \in {\mathbb{C}^{{N_t} \times {N_{RF}}}}:\left| {{{\mathbf{F}}_{PS}}(i,j)} \right| = 1,\forall i,j} \right\}$ and ${\mathcal{M}_{{{\mathbf{F}}_{BB}}}} = \left\{ {{{\mathbf{F}}_{BB}} \in {\mathbb{C}^{{N_{RF}} \times {N_S}}} : }\right.$ \\ $\left.{{{\left\| {{{\mathbf{F}}_{BB}}} \right\|}_F} = \sqrt {\frac{{{N_S}{N_{RF}}}}{{{N_t}}}} } \right\}$.
For the RPM, the tangent space ${T_{\left( {{{\mathbf{F}}_{PS}},{{\mathbf{F}}_{BB}}} \right)}}\mathcal{M}$ can be decomposed as a product of two tangent spaces
\begin{equation}
	\begin{aligned}
		&{T_{\left( {{{\bf{F}}_{PS}},{{\bf{F}}_{BB}}} \right)}}{\mathcal M}{\text{  =  }}{T_{{{\bf{F}}_{PS}}}}{{\mathcal M}_{{{\bf{F}}_{PS}}}} \times {T_{{{\bf{F}}_{BB}}}}{{\mathcal M}_{{{\bf{F}}_{BB}}}} \\
		&{\text{ =  }}\left\{ {\left( {{{\bm{\zeta }}_{{{\bf{F}}_{PS}}}},{{\bm{\zeta }}_{BB}}} \right)} \right.:{{\bm{\zeta }}_{{{\bf{F}}_{PS}}}} \in {{\mathbb C}^{{N_t} \times {N_{RF}}}},{{\bm{\zeta }}_{{{\bf{F}}_{BB}}}} \in {{\mathbb C}^{{N_{RF}} \times {N_S}}}, \\
		&\left.\qquad {\mathfrak{Re} \left( {{{\bm{\zeta }}_{{{\bf{F}}_{PS}}}} \circ {\bf{F}}_{PS}^*} \right) = {\bf{0}}}, \mathfrak{Re} \left( {{\text{tr}}\left( {{\bf{F}}_{BB}^H{{\bm{\zeta }}_{{{\bf{F}}_{BB}}}}} \right)} \right) = 0 \right\}
	\end{aligned}
\end{equation}

To facilitate the algorithm development, we need to define the Riemannian gradient over the product manifold, which is calculated as \cite{Mani1,Mani2,Mani3}
\begin{equation}
	\begin{aligned}
		& {\text{grad}}J\left( {{{\bf{F}}_{PS}},{{\bf{F}}_{BB}}} \right) \\
		& = \left( {{\text{gra}}{{\text{d}}_{{{\bf{F}}_{PS}}}}J\left( {{{\bf{F}}_{PS}},{{\bf{F}}_{BB}}} \right),{\text{gra}}{{\text{d}}_{{{\bf{F}}_{BB}}}}J\left( {{{\bf{F}}_{PS}},{{\bf{F}}_{BB}}} \right)} \right)
	\end{aligned}
\end{equation}
where the definition of ${\text{gra}}{{\text{d}}_{{{\mathbf{F}}_{PS}}}}J\left( {{{\mathbf{F}}_{PS}},{{\mathbf{F}}_{BB}}} \right)$ and ${\text{gra}}{{\text{d}}_{{{\mathbf{F}}_{BB}}}}J\left( {{{\mathbf{F}}_{PS}},{{\mathbf{F}}_{BB}}} \right)$ and detailed derivations are provided in Appendix \ref{app:1}.

Unlike the Riemannian manifold first-order optimization algorithm, for instance, RCG method in section \ref{MADMMalgo}, the second-order method adopts the Riemannian Hessian for quadratic information about the objective function, which enjoys more rapid convergence.
Thus, to utilize second-order algorithm solving \eqref{RPM1}, we need derive the explicit expression of the Riemannian Hessian in RPM.
The Riemannian Hessian of $J\left( {{{\bf{F}}_{PS}},{{\bf{F}}_{PS}}} \right)$ at $\left( {{{\bf{F}}_{PS}},{{\bf{F}}_{PS}}} \right) \in \mathcal{M}$ is a linear operator ${\text{Hes}}{{\text{s}}_{\left( {{{\bf{F}}_{PS}},{{\bf{F}}_{BB}}} \right)}}J:{T_{\left( {{{\bf{F}}_{PS}},{{\bf{F}}_{BB}}} \right)}}{\mathcal M}{\text{ }} \mapsto {T_{\left( {{{\bf{F}}_{PS}},{{\bf{F}}_{BB}}} \right)}}{\mathcal M}$ \cite{Mani1,Mani2,Mani3}, which are derived in Appendix \ref{app:2}.

\subsection{Riemannian Product Manifold Trust Region Algorithm}
Based on the above concepts, below we develop the RPM-TR algorithm \cite{Mani1,Mani2,TR2,Mani4} to solve the problem. 
Trust region methods \cite{TR2} can be understood as an enhancement of Newton’s method.
Based on \cite{Mani1,Mani2}, 
the main idea of RPM-TR can be summarized in three steps:
1) Generate a model that is a good approximation of the objective function in a specific region,
2) find the step that minimizes the model function in that specific region, determined by the trust region size,  
and 3) compute a new iterate based on a retracting mapping.

Firstly, we use a quadratic  function to approximate the  objective function \eqref{RPM1} at the $k$th iteration with the following form:
\begin{equation}
	\begin{aligned}
		& {w_k}\left( {{{\bm{\zeta}}_{\left( {{\bf{F}}_{PS}^{(k)},{\bf{F}}_{BB}^{(k)}} \right)}}} \right)  \\
		& =    J\left( {{\bf{F}}_{PS}^{(k)},{\bf{F}}_{BB}^{(k)}} \right) + \left\langle {{\text{grad}}J\left( {{\bf{F}}_{PS}^{(k)},{\bf{F}}_{BB}^{(k)}} \right),{{\bm{\zeta}}_{\left( {{\bf{F}}_{PS}^{(k)},{\bf{F}}_{BB}^{(k)}} \right)}}} \right\rangle   \\ 
		& \;\;  + \frac{1}{2}\left\langle {{\text{Hess}}J\left( {{\bf{F}}_{PS}^{(k)},{\bf{F}}_{BB}^{(k)}} \right)[{{\bm{\zeta}}_{\left( {{\bf{F}}_{PS}^{(k)},{\bf{F}}_{BB}^{(k)}} \right)}}],{{\bm{\zeta}}_{\left( {{\bf{F}}_{PS}^{(k)},{\bf{F}}_{BB}^{(k)}} \right)}}} \right\rangle  \cr
	\end{aligned}
	\label{TR_Model}
\end{equation}
Furthermore, since the model is only a local approximation of the pullback, we only trust it in a ball around the origin in the tangent space.
Thus, in the RPM-TR framework,  the update of $ \left( {{{\mathbf{F}}_{RF}},{{\mathbf{F}}_{BB}}} \right) $ within the trust region radius ${\Delta _k}$ is attained by solving 
\begin{equation}
	\begin{aligned}
		\mathop {{\text{min}}}\limits_{\left( {{{\bf{F}}_{RF}},{{\bf{F}}_{BB}}} \right)} & {w_k}\left( {{{\bm{\zeta}}_{\left( {{{\bf{F}}_{RF}},{{\bf{F}}_{BB}}} \right)}}} \right) \\
		{\text{s}}{\text{.t}}{\text{.}} \;\quad & \left\langle {{{\bm{\zeta}}_{\left( {{{\bf{F}}_{RF}},{{\bf{F}}_{BB}}} \right)}},{{\bm{\zeta}}_{\left( {{{\bf{F}}_{RF}},{{\bf{F}}_{BB}}} \right)}}} \right\rangle \leqslant \Delta _k^2 
	\end{aligned}
	\label{RPM2}
\end{equation}

Next, an approximate solution ${{{\bm{\zeta}}_{\left( {{\bf{F}}_{PS}^{(k)},{\bf{F}}_{BB}^{(k)}} \right)}}}$ of the trust region subproblem \eqref{RPM2} is computed, for example using a interior point method.
Notice that the approximate solution is a local minimum point in tangent space, i.e. ${{\bm{\zeta}}_{\left( {{\mathbf{F}}_{PS}^{(k)},{\mathbf{F}}_{BB}^{(k)}} \right)}} \in {T_{\left( {{\mathbf{F}}_{PS}^{(k)},{\mathbf{F}}_{BB}^{(k)}} \right)}}\mathcal{M}$.

Therefore, in the last step, the candidate for the new iterate is obtained by retraction mapping the approximate solution to the product manifold $\mathcal{M}$, which is
\begin{equation}
	\left( {{\bf{F}}_{PS}^{(k)\diamondsuit },{\bf{F}}_{BB}^{(k)\diamondsuit }} \right) = {{\mathcal R}_{\left( {{\bf{F}}_{PS}^{(k)},{\bf{F}}_{BB}^{(k)}} \right)}}\left( {{{\bm{\zeta}}_{\left( {{\bf{F}}_{PS}^{(k)},{\bf{F}}_{BB}^{(k)}} \right)}}} \right)
	\label{PRMstep1}
\end{equation}
where 
\begin{equation}
	{\begin{gathered}
			{\mathcal{R}_{\left( {{{\mathbf{F}}_{PS}},{{\mathbf{F}}_{BB}}} \right)}}\left( {{{\bm{\zeta}}_{\left( {{{\mathbf{F}}_{PS}},{{\mathbf{F}}_{BB}}} \right)}}} \right) 
			= \left( {{\mathcal{R}_{{{\mathbf{F}}_{PS}}}}\left( {{{\bm{\zeta}}_{{{\mathbf{F}}_{PS}}}}} \right),{\mathcal{R}_{{{\mathbf{F}}_{BB}}}}\left( {{{\bm{\zeta}}_{{{\mathbf{F}}_{BB}}}}} \right)} \right) \hfill \\ 
	\end{gathered}}
\end{equation}
with  ${{\mathcal{R}_{{{\mathbf{F}}_{PS}}}}\left( {{{\bm{\zeta}}_{{{\mathbf{F}}_{PS}}}}} \right)}$ and ${\mathcal{R}_{{{\mathbf{F}}_{BB}}}}\left( {{{\bm{\zeta}}_{{{\mathbf{F}}_{BB}}}}} \right)$ being retraction operators on manifold ${\mathcal{M}_{{{\mathbf{F}}_{PS}}}}$ and ${\mathcal{M}_{{{\mathbf{F}}_{BB}}}}$, respectively, and being defined as
\begin{subequations}
	\begin{equation}
		{{\mathcal R}_{{{\bf{F}}_{PS}}}}\left( {{\bm{\zeta} _{{{\bf{F}}_{PS}}}}} \right) = \left( {{{\bf{F}}_{PS}}{\text{  +  }}{\bm{\zeta} _{{{\bf{F}}_{PS}}}}} \right) \circ \frac{{\text{1}}}{{\left| {\left( {{{\bf{F}}_{PS}}{\text{  +  }}{\bm{\zeta} _{{{\bf{F}}_{PS}}}}} \right)} \right|}}
	\end{equation}
	\begin{equation}
		{{\mathcal{R}_{{{\mathbf{F}}_{BB}}}}\left( {{{\bm{\zeta}}_{{{\mathbf{F}}_{BB}}}}} \right) = \sqrt {\frac{{{N_S}{N_{RF}}}}{{{N_t}}}} \frac{{\left( {{{\mathbf{F}}_{BB}} + {{\bm{\zeta}}_{{{\mathbf{F}}_{BB}}}}} \right)}}{{{{\left\| {{{\mathbf{F}}_{BB}} + {{\bm{\zeta}}_{{{\mathbf{F}}_{BB}}}}} \right\|}_F}}}}
	\end{equation}
\end{subequations}

Once  obtaining the candidate form \eqref{PRMstep1}, the quality of the model \eqref{TR_Model} is assessed by forming the quotient
\begin{equation}
	{\rho _k} = \frac{{J\left( {{\bf{F}}_{PS}^{(k)},{\bf{F}}_{BB}^{(k)}} \right) - J\left( {{\bf{F}}_{PS}^{(k)\diamondsuit },{\bf{F}}_{BB}^{(k)\diamondsuit }} \right)}}{{{w_k}\left( {{{\bf{0}}_{\left( {{\bf{F}}_{PS}^{(k)},{\bf{F}}_{BB}^{(k)}} \right)}}} \right) - {w_k}\left( {{{\bm{\zeta}}_{\left( {{\bf{F}}_{PS}^{(k)},{\bf{F}}_{BB}^{(k)}} \right)}}} \right)}}
	\label{PRMstep2}
\end{equation} 
Depending on the value of $\rho_k$, the candidate $\left( {{\mathbf{F}}_{PS}^{(k) \diamondsuit },{\mathbf{F}}_{BB}^{(k) \diamondsuit }} \right)$ will be accepted or discarded and the trust-region radius $ {\Delta _{k + 1}} $ will be updated \cite{Mani1,Mani2}.
Specifically, a specific procedure is accepted or rejected the $\left( {{\mathbf{F}}_{PS}^{(k) \diamondsuit },{\mathbf{F}}_{BB}^{(k) \diamondsuit }} \right)$ based on the rule \cite{Mani1,Mani2}:
\begin{equation}
	{\left( {{\mathbf{F}}_{PS}^{(k + 1)},{\mathbf{F}}_{BB}^{(k + 1)}} \right) = \left\{ \begin{gathered}
			\left( {{\mathbf{F}}_{PS}^{(k) \diamondsuit },{\mathbf{F}}_{BB}^{(k) \diamondsuit }} \right)\;\;{\text{if}}{\rho _k} > \rho '({\text{accept}}), \hfill \\
			\left( {{\mathbf{F}}_{PS}^{(k)},{\mathbf{F}}_{BB}^{(k)}} \right)\;\;\;\;\;\;{\text{otherwise(reject)}}{\text{.}} \hfill \\ 
		\end{gathered}  \right.}
	\label{PRMstep3}
\end{equation}

Finally, the trust region radius $ {\Delta _{k + 1}} $ needs to be updated in each iteration. More concretely, the following update of $ {\Delta _{k + 1}} $ is used \cite{Mani1,Mani2,Mani4}:
\begin{equation}
	{{\Delta _{k + 1}} = \left\{ \begin{gathered}
			\frac{1}{4}{\Delta _k}\;\;\;\;\;\;\;\;\;\;\;\;\;\;\;\;{\text{if}}{\rho _k} < \frac{1}{4} \hfill \\
			\min \left( {2{\Delta _k},\Delta '} \right)\;{\text{if}}{\rho _k} > \frac{3}{4}\;{\text{and}}\;\left\| {{{\bm{\zeta}}_{\left( {{\mathbf{F}}_{PS}^{(k)},{\mathbf{F}}_{BB}^{(k)}} \right)}}} \right\| = {\Delta _k} \hfill \\
			{\Delta _k}\;\;\;\;\;\;\;\;\;\;\;\;\;\;\;\;\;\;{\text{otherwise}} \hfill \\ 
		\end{gathered}  \right.}
	\label{PRMstep4}
\end{equation}

According to the above analysis, the proposed RPM-TR alogrithm for solving problem \eqref{RPM1} is summarized in Algorithm \ref{alg1}.

\begin{algorithm}[htb] 
	\caption{RPM-TR  alogrithm for partially-connected structure hybrid beamforming design} 
	\label{alg1} 
	\begin{algorithmic}[1] 
		\REQUIRE Initial iteration $\left( {{\mathbf{F}}_{PS}^{\left(0\right)},{\mathbf{F}}_{BB}^{\left(0\right)}} \right) \in \mathcal{M}$, maximum radius $\Delta ' > 0$, reference beamforming matrix ${{\bf{F}}_{Rad}}$, ${{\bf{F}}_{Com}}$, maximum iterations ${K_{\max }}$, gradient threshold $\delta$.
		\ENSURE ${\mathbf{F}}_{PS}^{\left(opt\right)},{\mathbf{F}}_{BB}^{\left(opt\right)}$.
		\STATE \textbf{Initialization:} Set $k=1$, ${\Delta _0} \in \left( {0,\Delta '} \right)$ and $\rho ' \in \left[ {0,\frac{1}{4}} \right)$.
		\FOR{$k \leqslant {K_{\max }}$ and ${\left\| {{\text{grad}}J\left( {{{\mathbf{F}}_{PS}},{{\mathbf{F}}_{BB}}} \right)} \right\|_F} > \delta$}
		\STATE Approximately solve the subproblem \eqref{RPM2}, obtaining ${{\bm{\zeta}}_{\left( {{\mathbf{F}}_{PS}^{(k)},{\mathbf{F}}_{BB}^{(k)}} \right)}} \in {T_{\left( {{\mathbf{F}}_{PS}^{(k)},{\mathbf{F}}_{BB}^{(k)}} \right)}}\mathcal{M}$;
		\STATE Calculate the candidate for next iterate according to \eqref{PRMstep1};
		\STATE Compute the ratio of actual to model improvement using \eqref{PRMstep2};
		\STATE Accept or reject the tentative next iterate depending on the rule \eqref{PRMstep3};
		\STATE Update the trust region radius according to \eqref{PRMstep4};
		\STATE $k = k + 1$;
		\ENDFOR
		\STATE ${\mathbf{F}}_{PS}^{\left(opt\right)} =  {{\mathbf{F}}_{PS}^{(k)}} $ and ${\mathbf{F}}_{BB}^{\left(opt\right)} = {\mathbf{F}}_{BB}^{(k)}$
	\end{algorithmic} 
\end{algorithm}

We end this section by analyzing the complexity of the proposed RPM-TR algorithm. 
For the RPM-TR algorithm, the computational complexity mainly consists of computations of Riemannian gradient and Riemannian Hessian with complexities of $\mathcal{O}\left( {N_t{N_{RF}}{N_s}} \right)$ and $\mathcal{O}\left( {N_tN_{RF}^2 + N_t{N_{RF}}{N_s}} \right)$, respectively.
Hence, the complexity of  the algorithm \ref{alg1} is ${\mathcal O}\left( {{I_3}\left( {N_t^2{N_{RF}} + N_t^2{N_{RF}}{N_s}} \right)} \right)$, where ${{I_{3}}}$ stands for the total iteration numbers of the algorithm.

\section{Simulation Results And Discussion}\label{Sim_Dis}

In this section, several sets of numerical simulations are presented to evaluate the performance of our proposed algorithms.
Unless otherwise specified, in all simulations, we adopt the parameters summarized in Table \ref{tab:1}.
For mmWave channel, the AoAs/DoAs of the all clusters $\phi _{il}^t,\phi _{il}^r$ are assumed to be uniformly distributed in $[0,2\pi)$.
As for the MADMM algorithm, we assume the maximum number of iterations is $N_{max}=200$.
The termination criterion of RPM-TR algorithm is the line-search returns a displacement vector smaller than $10^{-6}$.

For simplifying notation, we denote  the beamformers ${{\mathbf{F}}_{Com}}$ and ${{\mathbf{F}}_{Rad}}$  as `ZF precoder' and `Ref. beampattern', respectively.
For the fully-connected structure, the  BCD-based method proposed in \cite{BCD1,BCD2} is selected as the benchmark for comparison.
The alternating minimization (AltMin) algorithm \cite{HBI5} is also included to demonstrate the superiority of the proposed RPM-TR algorithm for the partial-connected structure.
We randomly set the same initial point for all methods to provide a fair comparison among different algorithms and structures.

\begin{table*}[]
	\centering
	\caption{Simulation Parameters}
	\begin{tabular}{cc||cc}
		\hline 
		Parameter                & Value         & Parameter            & Value         \\ \hline \hline
		No. of transmit antennas & $N_t = 32$    & No. of RF chains     & $N_{RF} = 16$ \\ \hline
		No. of receive antennas  & $N_r = 6$     & No. of data streams  & $N_s = 6$     \\ \hline
		No. of rays              & $N_{ray} = 5$ & No. of clusters      & $N_{ray}= 10$ \\ \hline
	\end{tabular}
	\label{tab:1}
\end{table*}

\subsection{Convergence Performance}
\begin{figure}[!tb]
	\centering
	\subfigure[]{
		\label{Sim_con-a}
		\includegraphics[height=0.32\linewidth]{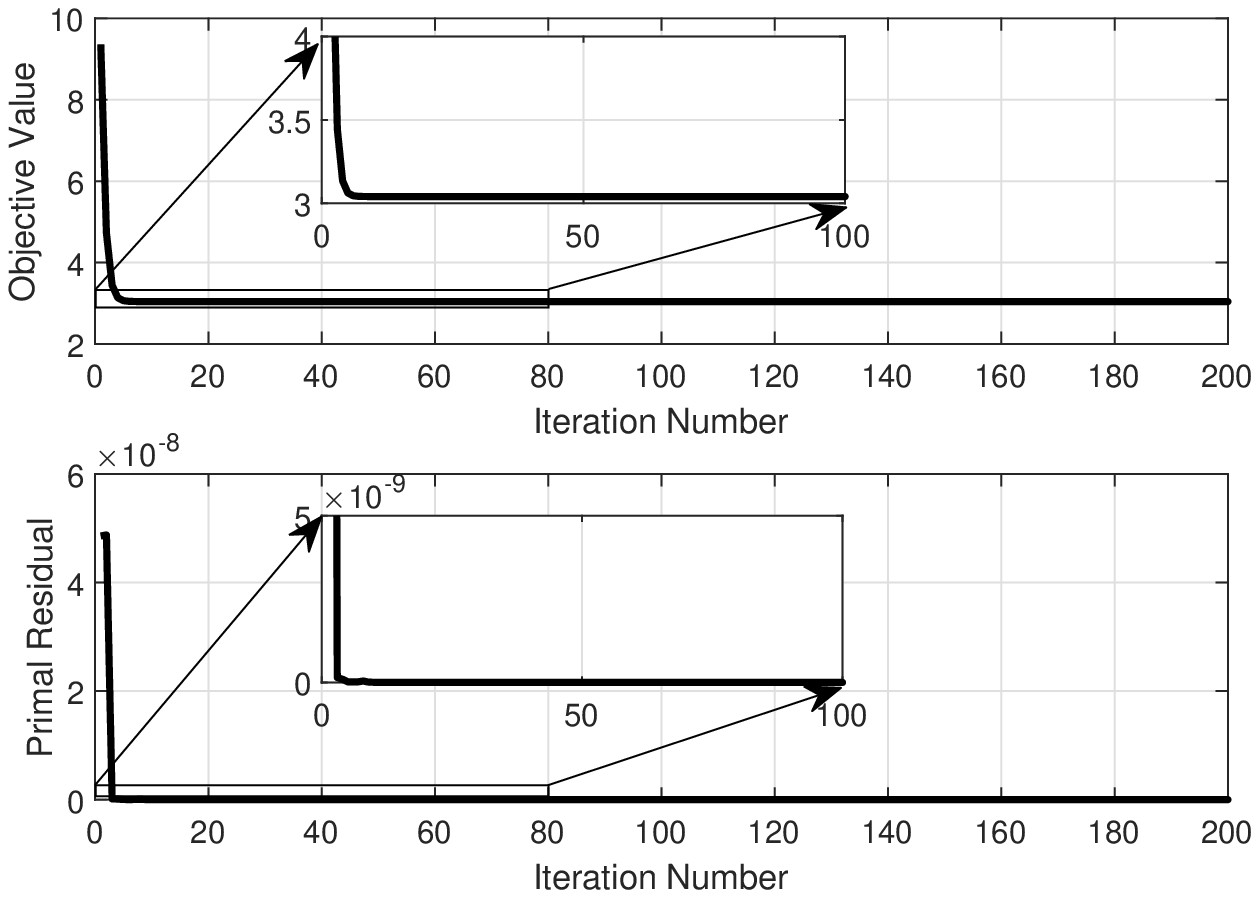}
	}	
	\subfigure[]{
		\label{Sim_con-b}
		\includegraphics[height=0.32\linewidth]{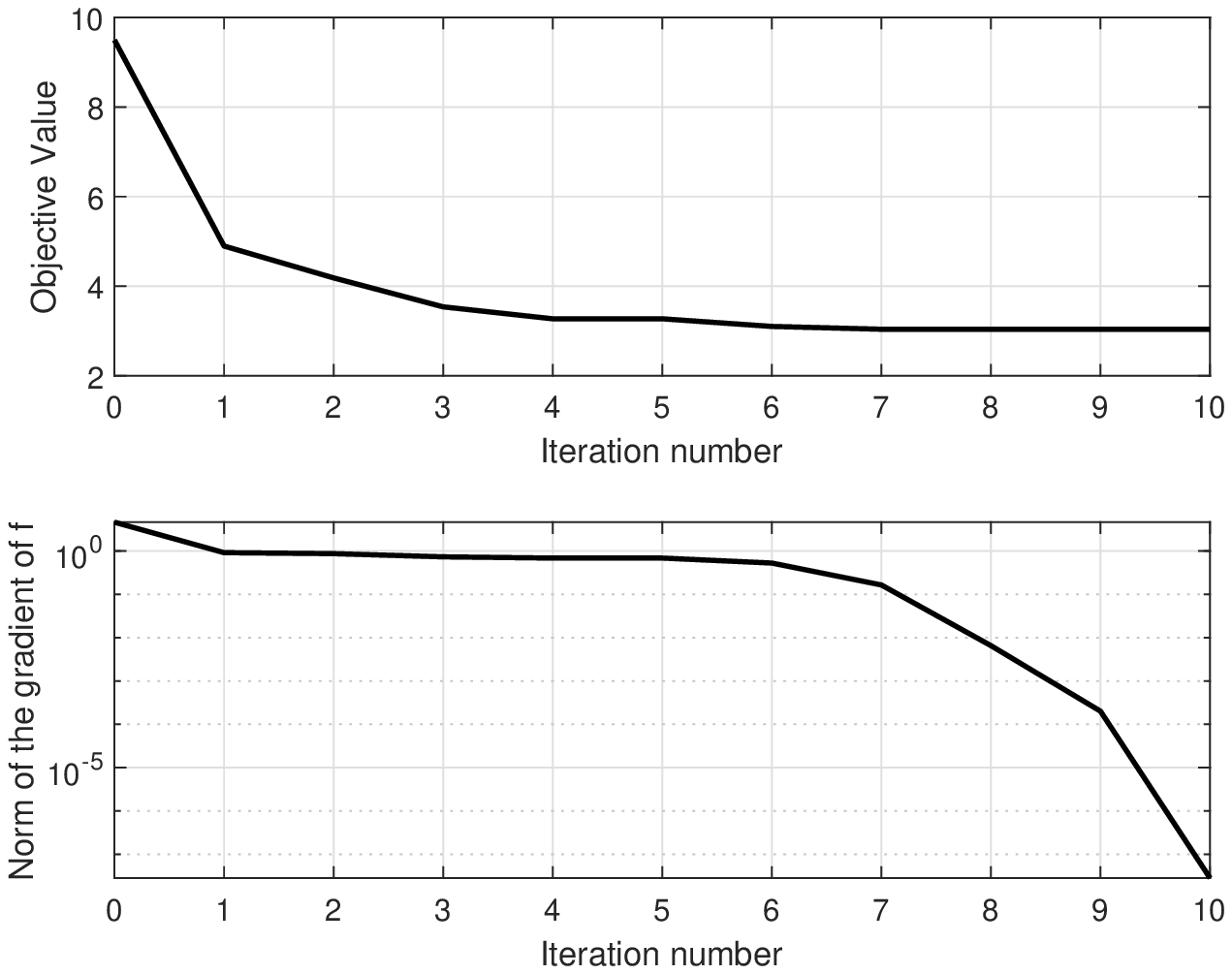}
	}
	\caption{The convergence performance of the RPM-TR algorithm and  MADMM algorithm. 
		(a) The convergence performance of the MADMM algorithm; (b) The convergence performance of the RPM-TR algorithm.
	}
	\label{Sim_con}
\end{figure}

We first show the resultant convergence performance of the proposed  MADMM and RPM-TR in Fig.\ref{Sim_con}.
The trade-off parameter for the considered case is set as $\varphi=0.5$.
Fig.\ref{Sim_con-a} presents the convergence behavior of the MADMM algorithm, which shows that the objective value in \eqref{ADMM1} increases and converges at a stationary point very fast.
Furthermore, the primal residual tend to zero as the iterative process goes on, which verifies the effectiveness of the proposed  MADMM algorithm.
For the RPM-TR method, in Fig.\ref{Sim_con-b}, the top plot shows the objective function in \eqref{RPM1} versus the number of iterations, which illustrates that the objective value decreases with the iteration number.
In addition, the norm of the Riemannian gradient displayed in Fig.\ref{Sim_con-b} bottom tends to $10^{-6}$ within 10 iterations, which indicates the excellent convergence performance of the RPM-TR algorithm.

\subsection{Spectral Efficiency Performance}

\begin{figure}[!tb]
	\centering
	\subfigure[]{
		\label{Sim_spec-a}
		\includegraphics[width=0.5\linewidth]{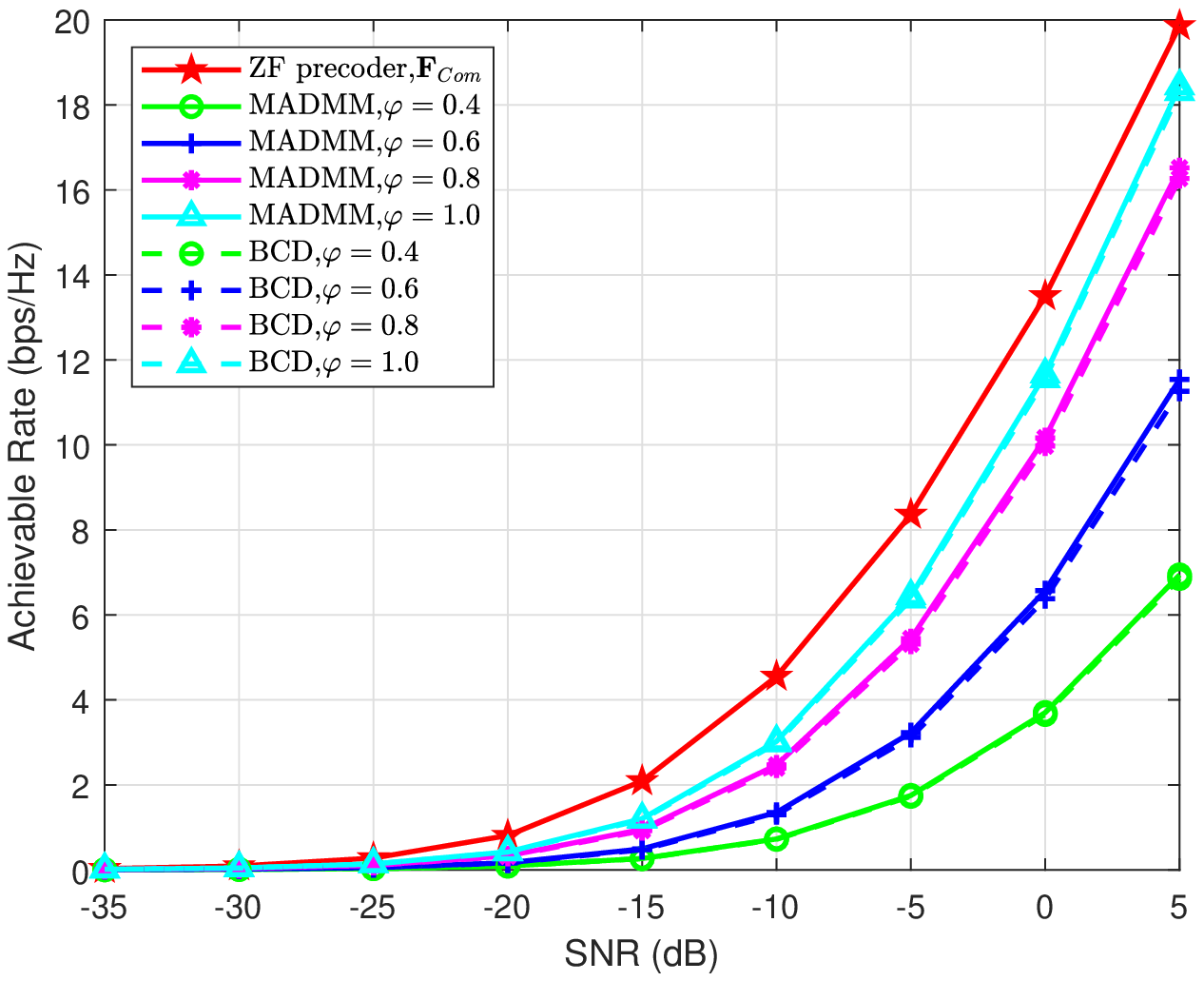}
	}
	\hspace{-2.5em}
	\subfigure[]{
		\label{Sim_spec-b}
		\includegraphics[width=0.5\linewidth]{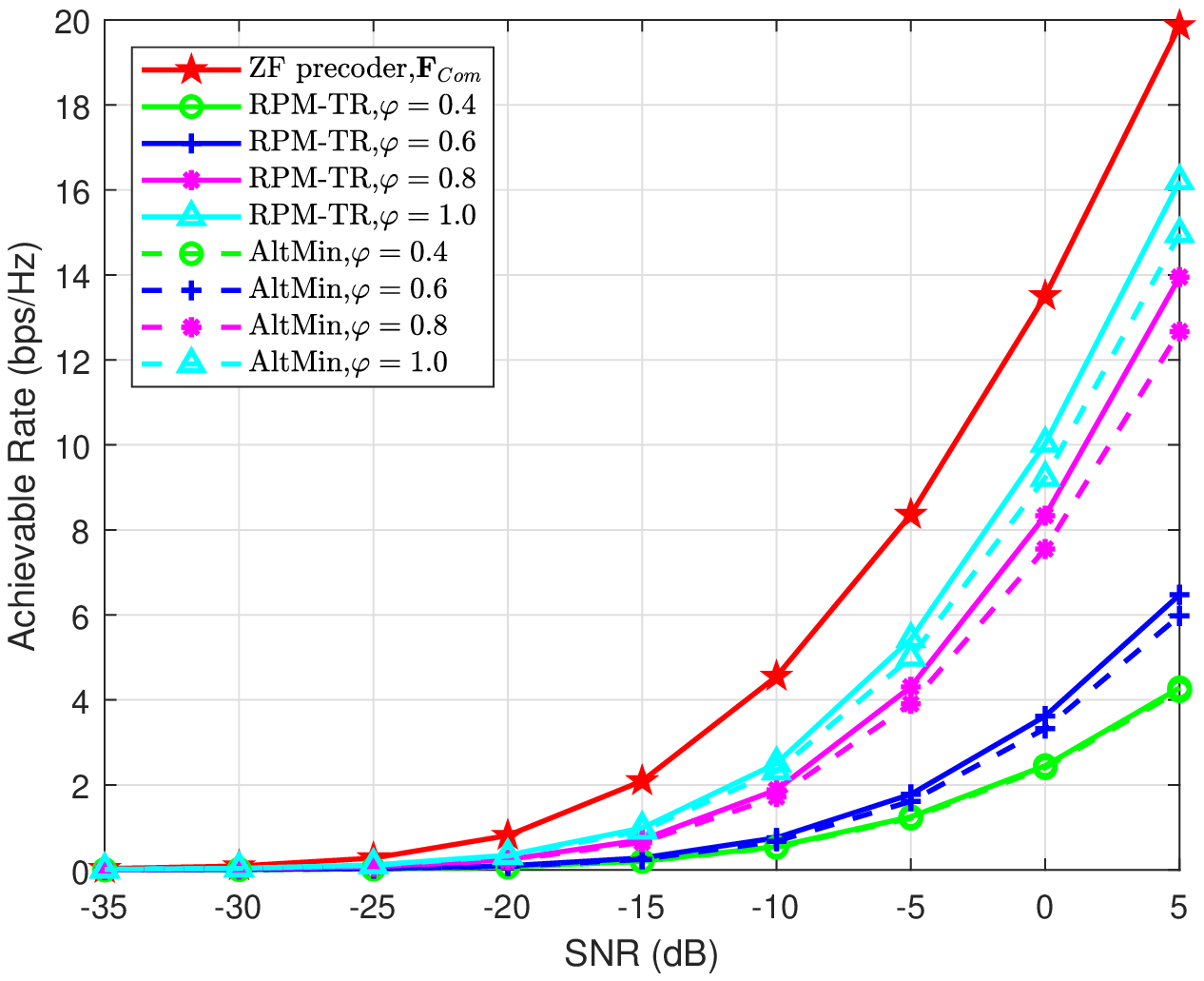}
	}
	\caption{The achievable rate values versus noise power for different trade-off parameter $\varphi$. (a) The achievable rate of the fully-connected structure  obtained by MADMM; (b) The achievable rate of the partially-connected structure  obtained by RPM-TR}
	\label{Sim_spec}
\end{figure}

In this subsection, we evaluate the downlink communication performance obtained by different algorithms.
Fig.\ref{Sim_spec-a} shows the achievable rate as a function of the SNR value and  trade-off parameter $\varphi$  for the fully-connected structure.  
As expected, the optimal digital precoder can attain the best achievable rate. 
In addition, the MADMM and BCD achievable rates increase with the SNR value and trade-off parameter.
We also find that the  MADMM algorithm performs slightly better than the BCD-based method.  

Fig.\ref{Sim_spec-b} compares the achievable rate obtained by the RPM-TR method to that by the  exiting AltMin algorithm for the partially-connected structure. 
The result in this figure shows s similar phenomenon to that in Fig.\ref{Sim_spec-a}.   
Moreover, it can be observed that  the proposed RPM-TR algorithm outperforms over the AltMin algorithm in terms of the achievable rate.

Furthermore, the comparison between two hybrid beamforming structures show that the partially-structure results in a non-negligible performance loss when comparing with the fully-connected structure.

\subsection{Beampattern Performance}

\begin{figure}[!tb]
	\centering
	\subfigure[]{
		\label{Sim_BP-a}
		\includegraphics[width=0.5\linewidth]{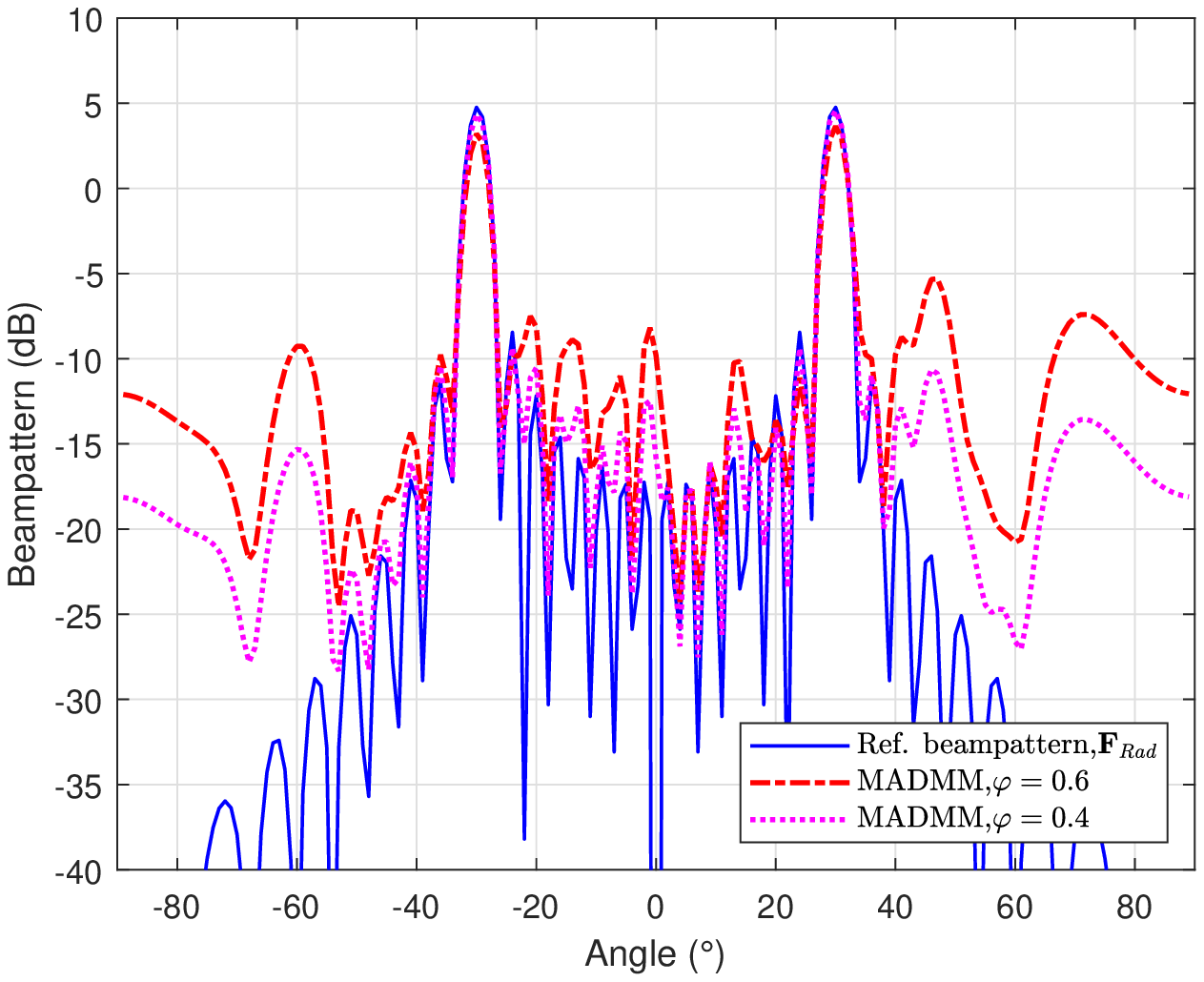}
	}
	\hspace{-2.5em}
	\subfigure[]{
		\label{Sim_BP-b}
		\includegraphics[width=0.5\linewidth]{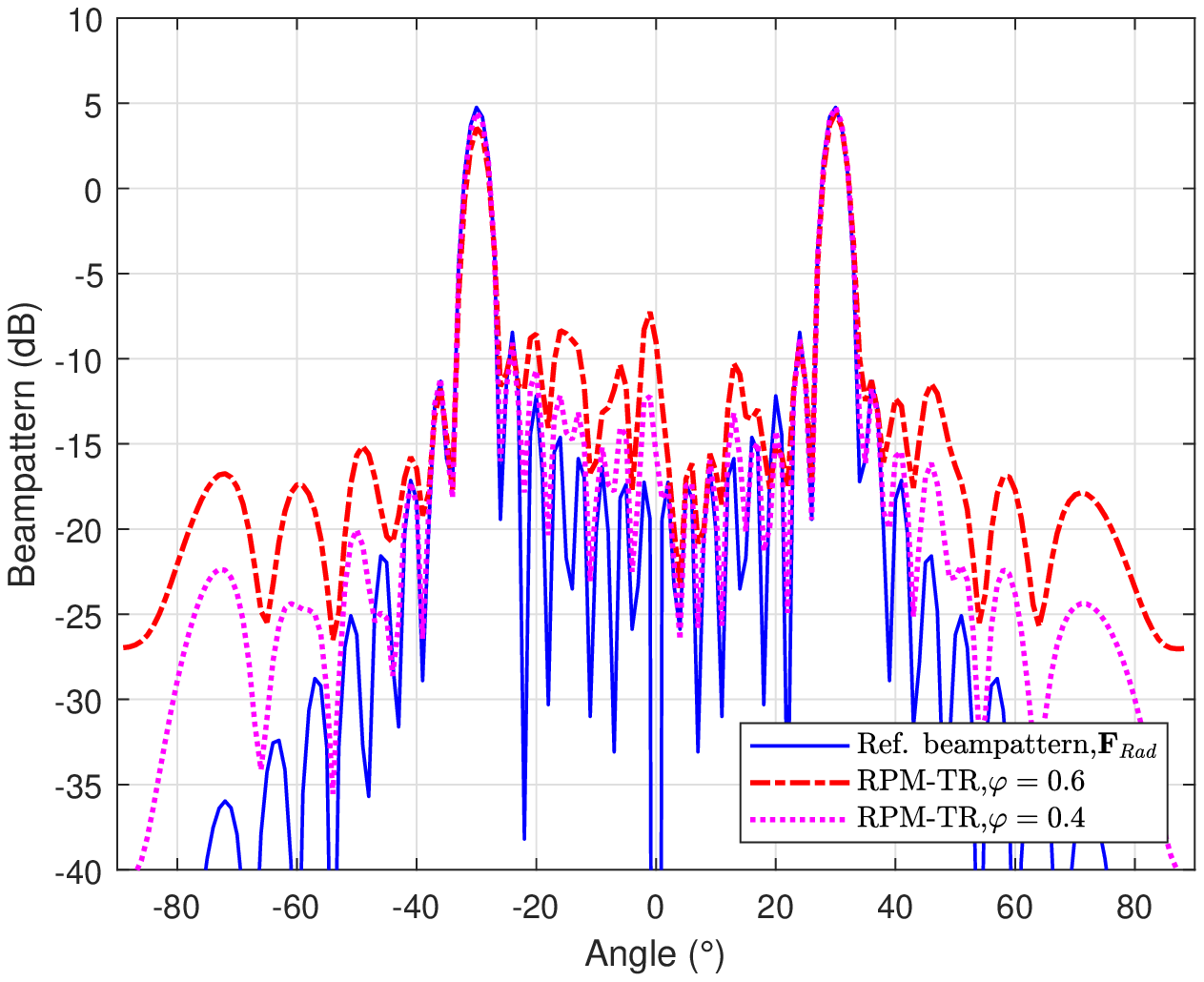}
	}
	\caption{Radar beampatterns obtained by RPM-TR and MADMM algorithm. (a) Radar beampatterns obtained by MADMM algorithm for fully-connected structure; (b) Radar beampatterns obtained by RPM-TR algorithm for partially-connected structure}
	\label{Sim_BP}
\end{figure}

In this subsection, we compare the performance of the achieved beampattern for different algorithms and structures.
We assume the targets locate at angle $\left\{-30^{\circ},30^{\circ}\right\}$.
Fig.\ref{Sim_BP} reveals that the smaller the trade-off parameter $\varphi$ is, i.e., the more critical weighting on the radar,  the better the sidelobe level of the designed beampattern. On the contrary, the achievable rates of downlink communication get the worse. 
As a result, a trade-off parameter should be delicately selected to achieve a good balance between communication and radar.

\begin{figure}[!tb]
	\centering
	\includegraphics[width=0.7\linewidth]{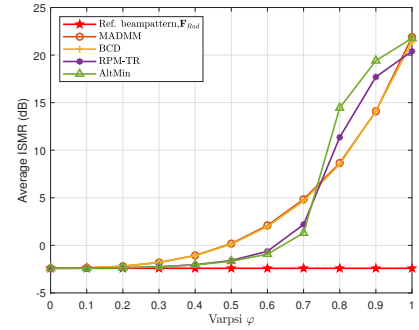}
	\caption{The ISMR of RPM-TR, AltMin, MADMM and BCD algorithms versus trade-off parameter}
	\label{Sim_ISMR}
\end{figure}

Fig. \ref{Sim_ISMR} shows the ISMR performance comparisons of the proposed algorithms with different analog structures.
It is seen from the figure that as $\varphi$ increases, the ISMR properties obtained by all algorithms become worse and worse. Moreover, it is interesting to observe that for  $\varphi\le 0.7$,  the algorithms of the partially-connected structure have ISMR behaviors than those of the fully-connected structure, while vice versa for   $\varphi \ge 0.8$.

\subsection{Performance With Different RF Chain numbers}

\begin{figure}[!tb]
	\centering
	\subfigure[]{
		\label{Sim_nrf-a}
		\includegraphics[width=0.5\linewidth]{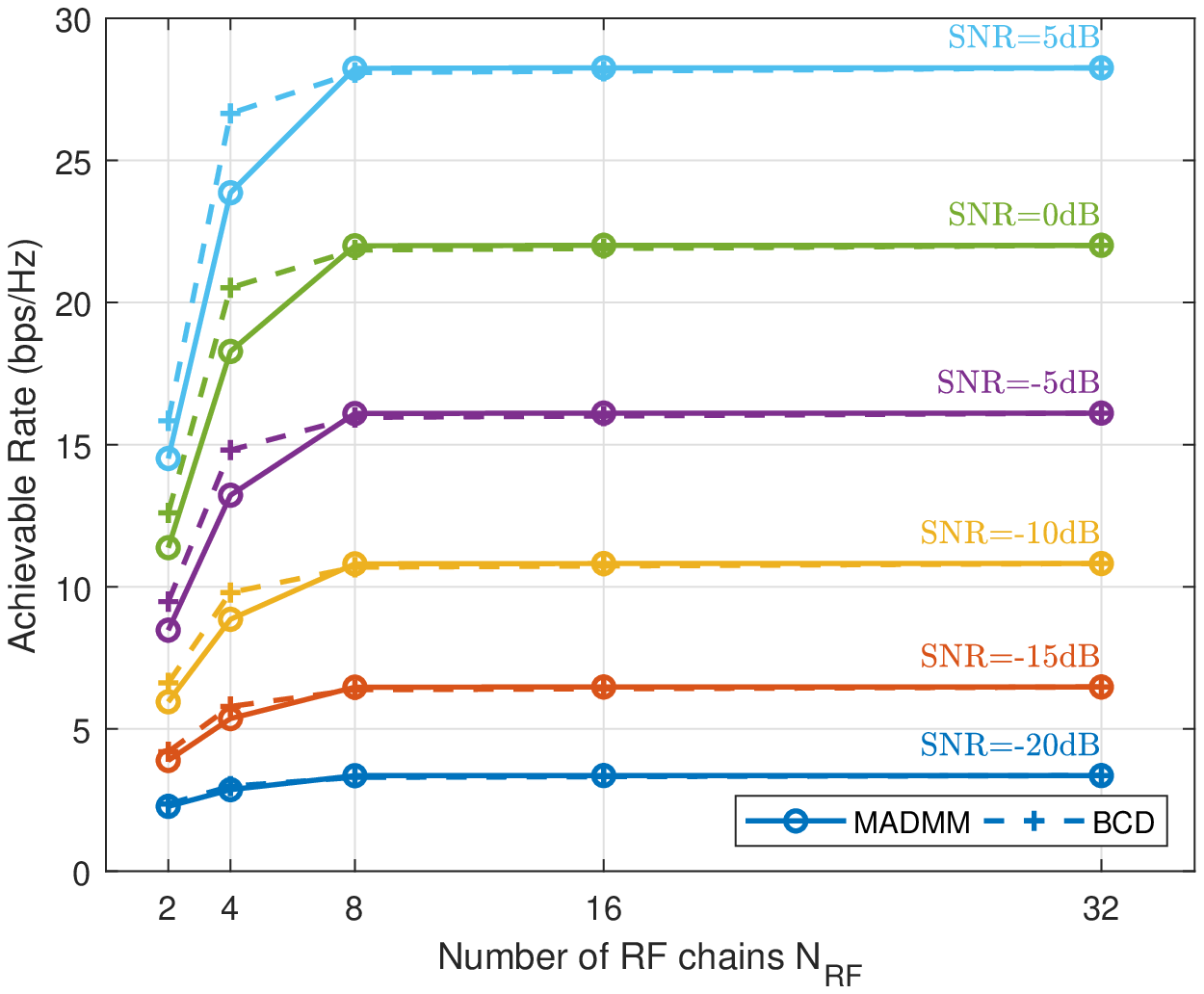}
	}
	\hspace{-2.5em}
	\subfigure[]{
		\label{Sim_nrf-b}
		\includegraphics[width=0.5\linewidth]{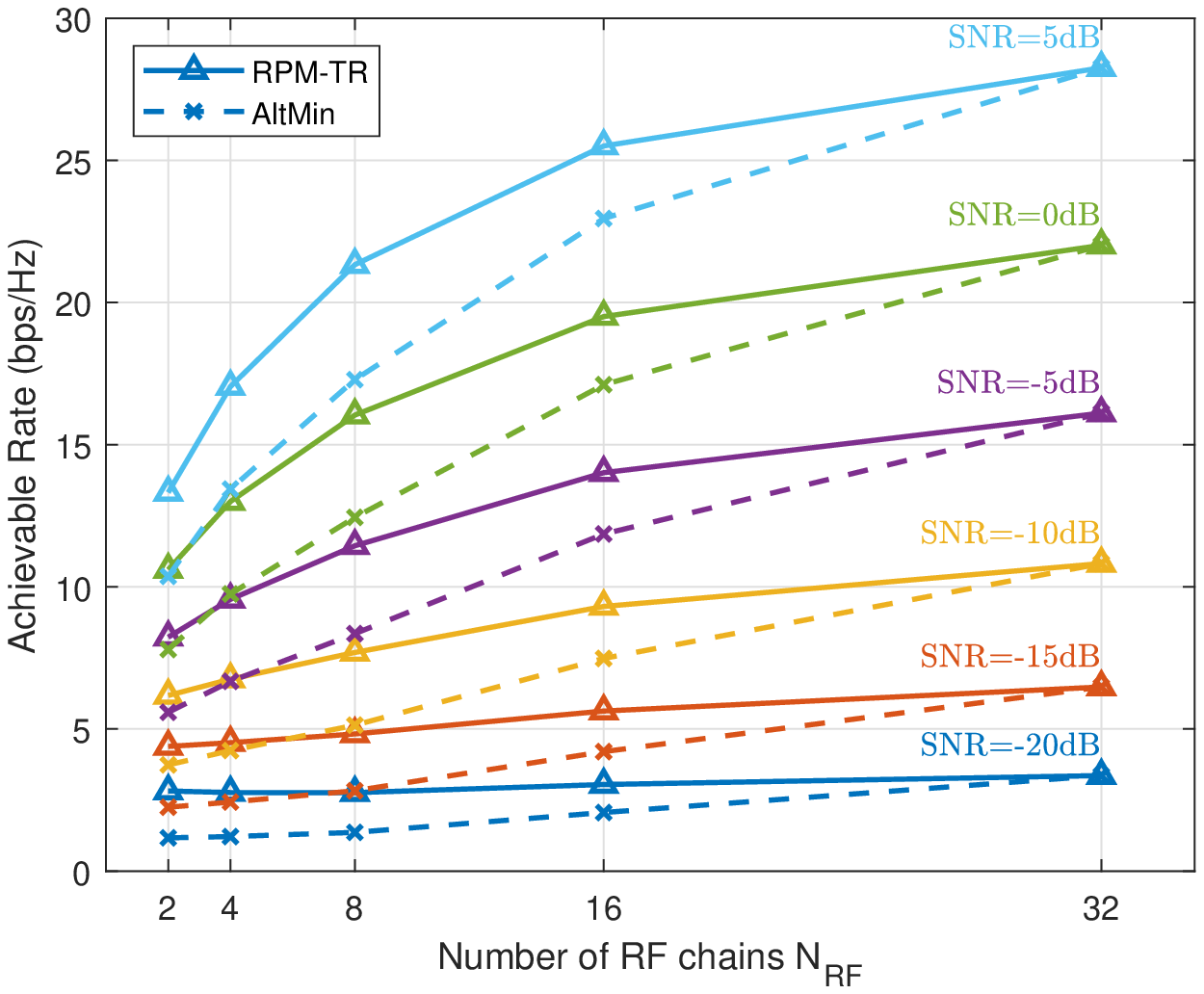}
	}
	\caption{The spectral efficiency values versus RF chains for different SNR with trade-off parameter $\varphi=0.4$. (a) Fully-connected structure, (b) Partially-connected structure. }
	\label{Sim_nrf}
\end{figure}

In this subsection, we will compare the performance of the two hybrid beamforming structures for different numbers of RF chains $N_{RF}$.
We assume the downlink user is equipped with $N_{r}=4$ antennas, which is equal to the number of data streams, i.e., $N_{S}=N_{r}=4$.

Fig.\ref{Sim_nrf-a} shows the spectral efficiency achieved by the MADMM method as a function of  the number of the SNR and $N_{RF}$ for the fully-connected structure. 
From the figure, we can see that although the proposed MADMM algorithm is worse than the BCD method when $N_{RF}<8$, the two methods provide nearly the same results when  $N_{RF}\ge 8$. 
Moreover, we also find that the achievable rates increase with the number of RF chains when $N_{RF} \le 8$, while the achievable rates are the same and equal to an upper bound when $N_{RF} > 8$. 
This indicates that it is possible to reduce the number of RF chains in the DFRC system to reduce the hardware cost.

Fig.\ref{Sim_nrf-b} displays the achievable spectral efficiency of the proposed hybrid beamforming method for partially-connected structure with different SNR. As expected, for the same SNR, the achievable spectral efficiency increase along with the $N_{RF}$.
Besides, the proposed RPM-TR algorithm consistently outperforms the AltMin method, and the gag becomes smaller with the increasing $N_{RF}$.

Combing the results in Fig.\ref{Sim_nrf-a} and Fig.\ref{Sim_nrf-b}, we note that the partially-connected structure is more sensitive to the number of $N_{RF}$ compared with a fully-connected one. 
Thus, choosing an appropriate number of RF chains in practice is essential to balance the performance and the cost.

\begin{table*}[]
	\centering
	\caption{Comparisons between different hybrid beamformers and methods}
	\begin{tabular}{ccccc}
		\hline
		Structure                            & \begin{tabular}[c]{@{}c@{}}Design\\ approach\end{tabular} & \begin{tabular}[c]{@{}c@{}}No. of \\ PS \end{tabular} & \begin{tabular}[c]{@{}c@{}}Computational \\ complexity \end{tabular} & Performance \\ \hline
		\multirow{2}{*}{Fully-connected}      &MADMM                                                      
		& $N_TN_{RF}$                                                                       
		& \small${\mathcal O}\left( {{I_0}{I_1}{N_s}{N_{RF}}{N_t}} \right)$                                                                    
		& \checkmark \checkmark \checkmark \checkmark            \\ \cline{2-5} 
		& BCD                                                       
		& $N_T N_{RF}$                                                                       
		& \small${\mathcal O}\left( {I_4}\left({{N_s}N_{RF}^3{N_t} + {N_s}N_{RF}^2N_t^2}\right) \right)$                                                                   
		& \checkmark \checkmark \checkmark \checkmark            \\ \hline
		\multirow{2}{*}{Partially-connected} 
		& RPM-TR                                                    
		& $N_{T}$                                                                           
		& \small${\mathcal O}\left( {{I_3}\left( {N_t^2{N_{RF}} + N_t^2{N_{RF}}{N_s}} \right)} \right)$                                                                   
		& \checkmark \checkmark \checkmark            \\ \cline{2-5} 
		& AltMin                                                   
		& $N_{T}$                                                                           
		& \small${\mathcal O}\left( {{I_5}\left( {N_t^2{N_s} + N_t^2{N_{RF}}} \right)} \right)$                                                                   
		& \checkmark \checkmark           \\ \hline
	\end{tabular}
	\label{tab:2}
\end{table*}

Finally, the comparisons between different hybrid beamformers and methods are listed in Table \ref{tab:2}. 
The results show that although the proposed MADMM is able to achieve nearly identical performance  to the  BCD approach for the fully-connected structure, it is more computationally efficient. 
For the partial-connected structure, the proposed RPM-TR outperforms the typical AltMin method  in terms of spectral efficiency, and they have similar complexity. 
For different structures, the fully-connected structure is benefit to improving the DFRC system performance while suffering from a higher hardware complexity.

\section{Conclusion}\label{Conlusion}
This paper addresses the problems of the hybrid beamforming designs for mmWave DFRC systems with two hybrid beamforming architectures, i.e., the partially-connected and fully-connected structures. The corresponding problems are formulated by the trade-off optimization criterion.
To deal with the non-convex formulated problems, we have proposed two low-complexity algorithms, concretely, the MADMM for fully-connected and the RPM-TR for the partially-connected structure, respectively. 
		
The simulation results reveal that the proposed methods can achieve satisfactory radar and communication performance for the DFRC with both fully-connected and partially-connected structures.
Besides, the proposed beamforming design provides a flexible trade-off between the radar and communication performance by adjusting a trade-off parameter.

Nevertheless, this initial work only investigates the DFRC system design with manifold optimization, and the particle implementation of these algorithms is still an open problem.
Moreover, based on this initial work, there are many issues worth being studied for future research, e.g., the scenarios for radar sensing in the presence of signal-dependent clutters, beam squint aware wideband system, as well as extending the proposed manifold based framework in different scenarios, e.g., reconfigurable intelligent surface aid DFRC.

\appendices

	\section{Derivation of Riemannian gradient}\label{app:1}
	Recalling the Riemannian gradient is calculated as
	\begin{equation}
		\begin{aligned}
			& {\text{grad}}J\left( {{{\bf{F}}_{PS}},{{\bf{F}}_{BB}}} \right) \\
			& = \left( {{\text{gra}}{{\text{d}}_{{{\bf{F}}_{PS}}}}J\left( {{{\bf{F}}_{PS}},{{\bf{F}}_{BB}}} \right),{\text{gra}}{{\text{d}}_{{{\bf{F}}_{BB}}}}J\left( {{{\bf{F}}_{PS}},{{\bf{F}}_{BB}}} \right)} \right)
		\end{aligned}
	\end{equation}
	where the Riemannian gradients ${\text{gra}}{{\text{d}}_{{{\mathbf{F}}_{PS}}}}J\left( {{{\mathbf{F}}_{PS}},{{\mathbf{F}}_{BB}}} \right)$ and ${\text{gra}}{{\text{d}}_{{{\mathbf{F}}_{BB}}}}J\left( {{{\mathbf{F}}_{PS}},{{\mathbf{F}}_{BB}}} \right)$ are computed by orthogonal projection from Euclidean space to the tangent space \cite{Mani2} as
	\begin{subequations}
		\begin{align}
			{\text{gra}} & {{\text{d}}_{{{\bf{F}}_{PS}}}}J\left( {{{\bf{F}}_{PS}},{{\bf{F}}_{BB}}} \right){\text{ }} \nonumber \\
			&= {\text{Pro}}{{\text{j}}_{{{\bf{F}}_{PS}}}}\left( {{\text{Gra}}{{\text{d}}_{{{\bf{F}}_{PS}}}}J\left( {{{\bf{F}}_{PS}},{{\bf{F}}_{BB}}} \right)} \right) \nonumber\\
			&= {\text{Gra}}{{\text{d}}_{{{\bf{F}}_{PS}}}}J\left( {{{\bf{F}}_{PS}},{{\bf{F}}_{BB}}} \right) \nonumber \\
			& - \mathfrak{Re} \{ {\text{Gra}}{{\text{d}}_{{{\bf{F}}_{PS}}}}J\left( {{{\bf{F}}_{PS}},{{\bf{F}}_{BB}}} \right) \circ {\bf{F}}_{PS}^*\}  \circ {{\bf{F}}_{PS}}  \\
			{\text{gra}} & {{\text{d}}_{{{\bf{F}}_{BB}}}}J\left( {{{\bf{F}}_{PS}},{{\bf{F}}_{BB}}} \right){\text{ }} \nonumber \\
			&= {\text{Pro}}{{\text{j}}_{{{\bf{F}}_{BB}}}}\left( {{\text{Gra}}{{\text{d}}_{{{\bf{F}}_{BB}}}}J\left( {{{\bf{F}}_{PS}},{{\bf{F}}_{BB}}} \right)} \right) \nonumber\\
			&= {\text{Gra}}{{\text{d}}_{{{\bf{F}}_{BB}}}}J\left( {{{\bf{F}}_{PS}},{{\bf{F}}_{BB}}} \right) \nonumber \\
			& - \mathfrak{Re} \left\{ {{\text{tr}}\left( {{\bf{F}}_{BB}^H{\text{Gra}}{{\text{d}}_{{{\bf{F}}_{BB}}}}g\left( {{{\bf{F}}_{PS}},{{\bf{F}}_{BB}}} \right)} \right)} \right\}{{\bf{F}}_{BB}} 
		\end{align}
	\end{subequations}
	where  ${{\text{Gra}}{{\text{d}}_{{{\mathbf{F}}_{PS}}}}J\left( {{{\mathbf{F}}_{PS}}{\mathbf{,}}{{\mathbf{F}}_{BB}}} \right)}$ and ${{\text{Gra}}{{\text{d}}_{{{\mathbf{F}}_{BB}}}}J\left( {{{\mathbf{F}}_{PS}}{\mathbf{,}}{{\mathbf{F}}_{BB}}} \right)}$ denote, respectively, the Euclidean gradients with respect to ${{{\mathbf{F}}_{PS}}}$ and ${{{\mathbf{F}}_{BB}}}$, and they are calculated as
	\begin{subequations}
		\begin{align}
			{\text{Gra}} & {{\text{d}}_{{{\mathbf{F}}_{PS}}}}J\left( {{{\mathbf{F}}_{PS}},{{\mathbf{F}}_{BB}}} \right) \nonumber \\
			= & 2\varphi \left\{ {\left( {\left( {{{\bf{F}}_D} \circ {{\bf{F}}_{PS}}} \right){{\bf{F}}_{BB}} - {{\bf{F}}_{Com}}} \right){\bf{F}}_{BB}^H} \right\} \circ {{\bf{F}}_D} \nonumber\\
			&+ 2\left( {1 - \varphi } \right)\left\{ {\left( {\left( {{{\bf{F}}_D} \circ {{\bf{F}}_{PS}}} \right){{\bf{F}}_{BB}} - {{\bf{F}}_{Rad}}} \right){\bf{F}}_{BB}^H} \right\} \circ {{\bf{F}}_D} \label{egrad1a} \\ 
			{\text{Gra}} & {{\text{d}}_{{{\mathbf{F}}_{BB}}}}J\left( {{{\mathbf{F}}_{PS}},{{\mathbf{F}}_{BB}}} \right) \nonumber \\
			= & 2\varphi {\left( {{{\bf{F}}_D} \circ {{\bf{F}}_{PS}}} \right)^H}\left\{ {\left( {{{\bf{F}}_D} \circ {{\bf{F}}_{PS}}} \right){{\bf{F}}_{BB}} - {{\bf{F}}_{Com}}} \right\} \nonumber \\
			&+ 2(1 - \varphi ){\left( {{{\bf{F}}_D} \circ {{\bf{F}}_{PS}}} \right)^H}\left\{ {\left( {{{\bf{F}}_D} \circ {{\bf{F}}_{PS}}} \right){{\bf{F}}_{BB}} - {{\bf{F}}_{Rad}}} \right\} \label{egrad1b}
		\end{align}
	\end{subequations}

	\section{Derivation of Riemannian Hessian}\label{app:2}
	The Riemannian Hessian of $J\left( {{{\bf{F}}_{PS}},{{\bf{F}}_{PS}}} \right)$ at $\left( {{{\bf{F}}_{PS}},{{\bf{F}}_{PS}}} \right) \in \mathcal{M}$ is a linear operator ${\text{Hes}}{{\text{s}}_{\left( {{{\bf{F}}_{PS}},{{\bf{F}}_{BB}}} \right)}}J:{T_{\left( {{{\bf{F}}_{PS}},{{\bf{F}}_{BB}}} \right)}}{\mathcal M}{\text{ }} \mapsto {T_{\left( {{{\bf{F}}_{PS}},{{\bf{F}}_{BB}}} \right)}}{\mathcal M}$ \cite{Mani1,Mani2,Mani3} defined by
	\begin{equation}
		\begin{aligned}
			{\text{Hes}} & {{\text{s}}_{\left( {{{\bf{F}}_{PS}},{{\bf{F}}_{BB}}} \right)}}J\left( {{{\bf{F}}_{PS}},{{\bf{F}}_{PS}}} \right)\left[ {{{\bm{\zeta }}_{\left( {{{\bf{F}}_{PS}},{{\bf{F}}_{BB}}} \right)}}} \right] \\
			& = {\nabla _{{{\bm{\zeta }}_{\left( {{{\bf{F}}_{PS}},{{\bf{F}}_{BB}}} \right)}}}}{\text{grad}}J\left( {{{\bf{F}}_{PS}},{{\bf{F}}_{PS}}} \right)
		\end{aligned}
	\end{equation}
	for the point ${{\bm{\zeta }}_{\left( {{{\bf{F}}_{PS}},{{\bf{F}}_{BB}}} \right)}} \in {T_{\left( {{{\bf{F}}_{PS}},{{\bf{F}}_{BB}}} \right)}}{\mathcal M}$.
	$\nabla $ denotes the Levi-Civita connection of $\mathcal{M}$.
	Specifically, for the RPM, the Riemannian Hessian of $J\left( {{{\bf{F}}_{PS}},{{\bf{F}}_{PS}}} \right)$ holds that \cite{Mani2}
	\begin{equation}
		\begin{aligned}
			&{\text{Hes}}{{\text{s}}_{\left( {{{\bf{F}}_{PS}},{{\bf{F}}_{BB}}} \right)}}J\left( {{{\bf{F}}_{PS}},{{\bf{F}}_{PS}}} \right)\left[ {{{\bm{\zeta }}_{\left( {{{\bf{F}}_{PS}},{{\bf{F}}_{BB}}} \right)}}} \right] \\
			&  = \left( {{\text{Hes}}{{\text{s}}_{{{\bf{F}}_{PS}}}}J\left( {{{\bf{F}}_{PS}},{{\bf{F}}_{PS}}} \right)\left[ {{{\bm{\zeta }}_{\left( {{{\bf{F}}_{PS}},{{\bf{F}}_{BB}}} \right)}}} \right],} \right. \\
			&  \qquad \qquad \left. {{\text{Hes}}{{\text{s}}_{{{\bf{F}}_{BB}}}}J\left( {{{\bf{F}}_{PS}},{{\bf{F}}_{PS}}} \right)\left[ {{{\bm{\zeta }}_{\left( {{{\bf{F}}_{PS}},{{\bf{F}}_{BB}}} \right)}}} \right]} \right)
		\end{aligned}
	\end{equation}
	where ${{\bm{\zeta}}_{\left( {{{\bf{F}}_{PS}},{{\bf{F}}_{BB}}} \right)}}$ is tangent vector on the  tangent space, i.e. ${{\bm{\zeta}}_{\left( {{{\bf{F}}_{PS}},{{\bf{F}}_{BB}}} \right)}} \in {T_{\left( {{{\bf{F}}_{PS}},{{\bf{F}}_{BB}}} \right)}}{\mathcal{M}}$,
	${{\text{Hes}}{{\text{s}}_{{{\bf{F}}_{PS}}}}J\left( {{{\bf{F}}_{PS}},{{\bf{F}}_{PS}}} \right)}$ and ${{\text{Hes}}{{\text{s}}_{{{\bf{F}}_{BB}}}}J\left( {{{\bf{F}}_{PS}},{{\bf{F}}_{PS}}} \right)}$ denote, respectively, the Riemannian Hessian of $J\left( {{{\bf{F}}_{PS}},{{\bf{F}}_{PS}}} \right)$ with respect to ${{{\bf{F}}_{PS}}}$ and ${{{\bf{F}}_{PS}}}$.
	Based on the classical expression of the Levi-Civita connection on a Riemannian submanifold of a Euclidean space \cite{Mani1,Mani2,Mani3}, the Riemannian Hessian can be calculated via direction derivative in the embedding space followed by an orthogonal projection to the tangent space.
	Thus, the Riemannian Hessian ${{\text{Hes}}{{\text{s}}_{{{\bf{F}}_{PS}}}}J\left( {{{\bf{F}}_{PS}},{{\bf{F}}_{PS}}} \right)}$ and ${{\text{Hes}}{{\text{s}}_{{{\bf{F}}_{BB}}}}J\left( {{{\bf{F}}_{PS}},{{\bf{F}}_{PS}}} \right)}$ are computed as \cite{Mani4,Mani6}
	\begin{subequations}
		\begin{align}
			&{\text{Hes}}{{\text{s}}_{{{\bf{F}}_{PS}}}}J\left( {{{\bf{F}}_{PS}},{{\bf{F}}_{PS}}} \right)\left[ {{{\bm{\zeta}}_{\left( {{{\bf{F}}_{PS}},{{\bf{F}}_{BB}}} \right)}}} \right] \nonumber\\
			&= {\nabla _{{{\bm{\zeta }}_{{{\bf{F}}_{PS}}}}}}{\text{grad}}J\left( {{{\bf{F}}_{PS}},{{\bf{F}}_{PS}}} \right) \nonumber \\
			& = {\text{Pro}}{{\text{j}}_{{{\bf{F}}_{PS}}}}\left( {{\text{Dgra}}{{\text{d}}_{{{\bf{F}}_{PS}}}}J\left( {{{\bf{F}}_{PS}},{{\bf{F}}_{PS}}} \right)\left[ {{{\bm{\zeta }}_{\left( {{{\bf{F}}_{PS}},{{\bf{F}}_{BB}}} \right)}}} \right]} \right) \nonumber \\
			&= {\text{Pro}}{{\text{j}}_{{{\bf{F}}_{PS}}}}\left( {{\text{eHes}}{{\text{s}}_{{{\bf{F}}_{PS}}}}J - \mathfrak{Re} \left( {{{\bf{F}}_{PS}} \circ {{\left( {{\text{Gra}}{{\text{d}}_{{{\bf{F}}_{PS}}}}J} \right)}^*}} \right) \circ {{\bm{\zeta}}_{{{\bf{F}}_{PS}}}}} \right) \\
			&{\text{Hes}}{{\text{s}}_{{{\bf{F}}_{BB}}}}{J\left( {{{\bf{F}}_{PS}},{{\bf{F}}_{PS}}} \right)}\left[ {{{\bm{\zeta}}_{\left( {{{\bf{F}}_{PS}},{{\bf{F}}_{BB}}} \right)}}} \right] \nonumber\\
			& = {\nabla _{{{\bf{\zeta }}_{{{\bf{F}}_{BB}}}}}}{\text{grad}}J\left( {{{\bf{F}}_{PS}},{{\bf{F}}_{PS}}} \right) \nonumber \\
			&{\text{ = Pro}}{{\text{j}}_{{{\bf{F}}_{BB}}}}\left( {{\text{Dgra}}{{\text{d}}_{{{\bf{F}}_{BB}}}}J\left( {{{\bf{F}}_{PS}},{{\bf{F}}_{PS}}} \right)\left[ {{{\bm{\zeta }}_{\left( {{{\bf{F}}_{PS}},{{\bf{F}}_{BB}}} \right)}}} \right]} \right) \nonumber \\
			& = {\text{Pro}}{{\text{j}}_{{{\bf{F}}_{BB}}}}\left( {{\text{eHes}}{{\text{s}}_{{{\bf{F}}_{BB}}}}J} \right) + \mathfrak{Re} \left\{ {{\text{tr}}\left( {{\bf{F}}_{BB}^H{\text{Gra}}{{\text{d}}_{{{\bf{F}}_{BB}}}}J} \right)} \right\}{{\bm{\zeta }}_{{{\bf{F}}_{BB}}}}
		\end{align}
	\end{subequations}
	where ${{\text{Dgra}}{{\text{d}}_{{{\bf{F}}_{PS}}}}J\left( {{{\bf{F}}_{PS}},{{\bf{F}}_{PS}}} \right)\left[ {{{\bf{\zeta }}_{\left( {{{\bf{F}}_{PS}},{{\bf{F}}_{BB}}} \right)}}} \right]}$ and ${{\text{Dgra}}{{\text{d}}_{{{\bf{F}}_{BB}}}}J\left( {{{\bf{F}}_{PS}},{{\bf{F}}_{PS}}} \right)\left[ {{{\bf{\zeta }}_{\left( {{{\bf{F}}_{PS}},{{\bf{F}}_{BB}}} \right)}}} \right]}$ are the directional derivative of the Riemannian gradient ${{\text{gra}}{{\text{d}}_{{{\bf{F}}_{PS}}}}J\left( {{{\bf{F}}_{PS}},{{\bf{F}}_{PS}}} \right)}$ and ${{\text{gra}}{{\text{d}}_{{{\bf{F}}_{BB}}}}J\left( {{{\bf{F}}_{PS}},{{\bf{F}}_{PS}}} \right)}$,
	the Euclidean Hessian ${\text{eHes}}{{\text{s}}_{{{\bf{F}}_{PS}}}}J$ and ${\text{eHes}}{{\text{s}}_{{{\bf{F}}_{BB}}}}J$ denote, respectively, the  directional derivative of the Euclidean gradient \eqref{egrad1a} and \eqref{egrad1b}, which are computed as
	\begin{subequations}
		\begin{align}
			{\text{e}}&{{\text{Hess}}_{{{\bf{F}}_{PS}}}}J   \nonumber\\
			&= 2\varphi \left\{ {\left( {\left( {{{\bf{F}}_D} \circ {{\bf{F}}_{PS}}} \right){{\bf{F}}_{BB}} - {{\bf{F}}_{Com}}} \right){\bm{\zeta }}_{BB}^H} \right\} \circ {{\bf{F}}_D}  \nonumber\\
			&\quad + 2\left( {1 - \varphi } \right)\left\{ {\left( {\left( {{{\bf{F}}_D} \circ {{\bf{F}}_{PS}}} \right){{\bf{F}}_{BB}} - {{\bf{F}}_{Rad}}} \right){\bm{\zeta }}_{BB}^H} \right\} \circ {{\bf{F}}_D} \nonumber\\
			&\quad + 2\left\{ {\left( {\left( {{{\bf{F}}_D} \circ {{\bm{\zeta }}_{PS}}} \right){{\bf{F}}_{BB}} + \left( {{{\bf{F}}_D} \circ {{\bf{F}}_{PS}}} \right){{\bm{\zeta }}_{BB}}} \right){\bf{F}}_{BB}^H} \right\} \circ {{\bf{F}}_D} \nonumber \\
			{\text{e}}&{{\text{Hess}}_{{{\bf{F}}_{BB}}}}J  \nonumber\\
			& = 2\varphi {\left( {{{\bf{F}}_D} \circ {{\bm{\zeta }}_{PS}}} \right)^H}\left\{ {\left( {{{\bf{F}}_D} \circ {{\bf{F}}_{PS}}} \right){{\bf{F}}_{BB}} - {{\bf{F}}_{Com}}} \right\}\nonumber\\
			&\quad + 2(1 - \varphi ){\left( {{{\bf{F}}_D} \circ {{\bm{\zeta }}_{PS}}} \right)^H}\left\{ {\left( {{{\bf{F}}_D} \circ {{\bf{F}}_{PS}}} \right){{\bf{F}}_{BB}} - {{\bf{F}}_{Rad}}} \right\} \nonumber\\
			&\quad + 2{\left( {{{\bf{F}}_D} \circ {{\bf{F}}_{PS}}} \right)^H}\left\{ {\left( {{{\bf{F}}_D} \circ {{\bm{\zeta }}_{PS}}} \right){{\bf{F}}_{BB}} + \left( {{{\bf{F}}_D} \circ {{\bf{F}}_{PS}}} \right){{\bm{\zeta }}_{BB}}} \right\} \nonumber
		\end{align}
	\end{subequations}

\footnotesize
 \balance
\bibliographystyle{IEEEtran}
\bibliography{IEEEabrv,ref.bib}

\end{document}